\documentclass[sigconf]{acmart}

\renewcommand\footnotetextcopyrightpermission[1]{}  
\AtBeginDocument{%
  \providecommand\BibTeX{{%
    \normalfont B\kern-0.5em{\scshape i\kern-0.25em b}\kern-0.8em\TeX}}}

\usepackage{url} 
\usepackage{float}
\usepackage{mathrsfs}
\usepackage{graphicx}
\usepackage{enumerate}

\usepackage{caption, subcaption}
\usepackage{kbordermatrix}
\usepackage{color}
\usepackage{algorithm}
\usepackage{algorithmic}
\usepackage{tcolorbox}
\acmConference[]{}{}{}
\newtheorem{thm1}{\citet{BarbCand20}'s Theorem}
\newtheorem*{thm2}{Theorem 1}

\usepackage{natbib}
\usepackage{multirow} 

\begin{document}

\title[Nonparametric Bayesian Knockoff Generators]{Nonparametric Bayesian Knockoff Generators for Feature Selection Under Complex Data Structure}

\author{Michael J. Martens}
\email{mmartens@mcw.edu}
\affiliation{%
	\institution{Medical College of Wisconsin}
	\streetaddress{8701 Watertown Plank Rd.}
	\city{Milwaukee}
	\state{WI}
	\country{USA}
	\postcode{53226}
}

\author{Anjishnu Banerjee}
\email{abanerjee@mcw.edu}
\affiliation{%
	\institution{Medical College of Wisconsin}
	\streetaddress{8701 Watertown Plank Rd.}
	\city{Milwaukee}
	\state{WI}
	\country{USA}
	\postcode{53226}
}

\author{Xinran Qi}
\email{xiqi@coh.org}
\affiliation{%
	\institution{City of Hope}
	\streetaddress{1500 E Duarte Rd}
	\city{Duarte}
	\state{CA}
		\country{USA}
	\postcode{91010}}

\author{Yushu Shi}
\email{yus4011@med.cornell.edu}
\affiliation{%
	\institution{Weill Cornell Medicine}
	\city{New York}
	\state{NY}
		\country{USA}
	\postcode{10065}
}

\renewcommand{\shortauthors}{Martens, et al.}

\begin{abstract}
The recent proliferation of high-dimensional data, such as electronic health records and genetics data, offers new opportunities to find novel predictors of outcomes. Presented with a large set of candidate features, interest often lies in selecting the ones most likely to be predictive of an outcome for further study. Controlling the false discovery rate (FDR) at a specified level is often desired in evaluating these variables. Knockoff filtering is an innovative strategy for conducting FDR-controlled feature selection. This paper proposes a nonparametric Bayesian model for generating high-quality knockoff copies that can improve the accuracy of predictive feature identification for variables arising from complex distributions, which can be skewed, highly dispersed and/or a mixture of distributions. The proposed Gaussian Dirichlet process mixture (GDPM) knockoff model offers two main benefits: (a) in cases where the data originate from a Gaussian mixture, the GDPM does not require pre-specifying the number of mixture components, instead learning the number of components from the data themselves and avoiding its potential mis-specification; and (b) the GDPM can accurately model a broad range of non-Gaussian data-generating distributions, such as heavy-tailed distributions, a capability theoretically supported. This paper provides a detailed description for generating knockoff copies from a GDPM model via MCMC posterior sampling. Additionally, we provide a theoretical guarantee on the robustness of the knockoff procedure. Through simulations, the method is shown to identify important features with accurate FDR control and improved power over the popular second-order Gaussian knockoff generator. Furthermore, the model is compared with finite Gaussian mixture knockoff generator in FDR and power. The proposed technique is applied for detecting genes predictive of survival in ovarian cancer patients using data from The Cancer Genome Atlas (TCGA).
\end{abstract}






\maketitle

\section{INTRODUCTION}
\label{sec:intro}
The proliferation of high dimensional data during the past two decades offers new opportunities to identify novel predictors of outcomes. Yet, this abundance also presents formidable challenges to accurately detecting these factors. Among a large number of candidate features, a large proportion is usually expected to have a negligible effect on the outcome. Interest lies in selecting just the ones most likely to be predictive, so the goal is often to control the false discovery rate (FDR) \citep{BenjHoch95} at a prespecified level. Penalized regression models, such as the lasso \citep{Tibs96}, are often employed for feature selection. However, formally controlling the FDR is generally quite challenging for these procedures. 

Knockoff filtering is a cutting-edge strategy that can select features while maintaining FDR control. The rationale of this approach proceeds as follows. First, for each candidate feature $X_j$, a knockoff copy $\widetilde{X}_j$ is constructed such that the knockoff feature is pairwise exchangeable to the original one but is independent of the outcome. Second, the influences of the original and knockoff features on the outcome are evaluated using quantities called importance statistics, e.g., the magnitude of the regression coefficients in lasso model or the lasso penalty parameter value at which a feature enters the model;
the independence of each knockoff feature from the outcome allows it to serve as a negative control for its corresponding original feature when assessing the original feature's influence. Third, features are selected based on importance statistics such that the FDR is controlled at the specified level. An advantage of this approach is that FDR control is maintained for finite samples, avoiding reliance on asymptotic properties. Although this procedure is straightforward, the construction of suitable knockoff features is nontrivial. 

The original knockoff procedure \citep{BarbCand15}, called Fixed-X, constructs knockoffs deterministically from the observed data and is applicable only when the number of features does not exceed the sample size ($p \le n$). Later,
\citet{CandFan18} proposed to generate knockoffs randomly by modeling the distribution of the original features, an approach called Model-X.
\textcolor{black}{ When the true covariate distribution is unknown, a popular knockoff construction method is second-order model-X knockoff, which generates knockoffs from a Gaussian model whose mean and variance are estimated from the original features'. However, matching only the first two moments may not be sufficient for real-life applications with data from complex distributions with properties such as skewness, multimodality, or mixture components. Mixture distributions are commonly seen in real life, including EHR data from multiple treatment centers or genetics data from patients with various disease subtypes.} Previously, \citet{GimeGhor19} proposed using a finite mixture of Gaussian distributions to construct knockoff copies to overcome this limitation. This should provide more higher quality knockoffs because, with a sufficiently large number of mixture components, any continuous distribution may be estimated arbitrarily closely by a mixture of Gaussians. However, this model requires prespecification of the number of components, a parameter for which little information may be available a priori. A Dirichlet process mixture (DPM) of Gaussian distributions \citep{Lo84} avoids this requirement by automatically providing both the number of mixture components and their parameters based on the observed data, leading to its successful application in accurate density estimation for continuous multidimensional data. We propose to utilize a Gaussian DPM model to generate continuous knockoff features, offering flexible and accurate modeling that can fully capture the complexity of arbitrary data distributions. 


The remainder of the article is organized as follows. Section 2 proposes a nonparametric Dirichlet process mixture of Gaussians model for high dimensional features. A theoretical guarantee for the robustness of the knockoffs is provided in Section 3. The performance of the Bayesian knockoff generator is evaluated and compared with second order Gaussian knockoff and finite mixture knockoff methods through simulations in Section 4. Section 5 demonstrates the application of our proposed method on an ovarian cancer gene expression dataset. Section 6 concludes the paper with a brief discussion and lays our our future directions.

\section{GAUSSIAN DIRICHLET PROCESS MIXTURE KNOCKOFF GENERATOR}
This section introduces the Gaussian Dirichlet Process Mixture (GDPM) model for generating knockoffs and provides a prior specification that works in practice for fitting this model. The GDPM is a useful tool for modeling the potentially complex data distribution of high-dimensional data vectors. We utilize the GDPM model for knockoff generation in this context by estimating the distribution of the features $\mathbf{X}$ and then generating knockoff copies $\widetilde{\mathbf{X}}$ from the estimated distribution. 
\subsection{Model and Hyperparameter Specification}
A Gaussian distribution can be parameterized by its associated mean vector and precision matrix, denoted as the pair $\boldsymbol{\theta} = (\boldsymbol{\mu},\boldsymbol{\Omega})$ taking values in the set $\{(\mathbf{a},\mathbf{B}) : \mathbf{a} \in \mathbb{R}^p, \mathbf{B} \in \mathbb{R}^{p \times p} \text{ is positive definite}\}$. The GDPM model specifies that the rows $\mathbf{X}_i$ of the feature matrix $\mathbf{X}_{n\times p}$ arise from an infinite mixture of these distributions, specified as follows:
\begin{equation*}
	\label{GDPMmodel}
	\begin{split}
		\mathbf{X}_i  | \boldsymbol{\theta}_i &\overset{iid}\sim \mathrm{No}(\boldsymbol{\mu}_i, \boldsymbol{\Omega}_i), \quad 1 \le i \le n;\\
		\qquad (\boldsymbol{\mu}_i, \boldsymbol{\Omega}_i) & \overset{ind}\sim G, \quad 1 \le i \le n; \\
		G &\sim \mathrm{DP}(G_0, \alpha),\\
		\quad G_0 &= \mathrm{NoWi}(\mathbf{m}, k, \mathbf{W}, v),\\
		\alpha&\sim \mathrm{Ga}(a,b).
	\end{split}
\end{equation*}
$\mathrm{DP}(G_0,\alpha)$ is a Dirichlet process with base distribution $G_0$ and concentration parameter $\alpha$. $G_0$ is given a Normal-Wishart distribution, which is conjugate to the Gaussian data distribution. To be precise, $(\mathbf{u}, \mathbf{V}) \sim \mathrm{NoWi}(\mathbf{m}, k, \mathbf{W}, v)$ means that $\mathbf{V} \sim \mathrm{Wishart}(\mathbf{W},v)$ has scale matrix $\mathbf{W}$, degrees of freedom $v$ and $\mathbf{u} | \mathbf{V} \sim \mathrm{No}(\mathbf{m},k\mathbf{V})$ has mean $\mathbf{m}$, precision matrix $k\mathbf{V}$. The concentration parameter $\alpha$ controls how much variability arises in the distributions generated by the Dirichlet process, with higher $\alpha$ giving less variability and distributions that are more similar to $G_0$. The $G$ distribution generated by the Dirichlet process is discrete, which tends to cause some $\boldsymbol{\theta}_i = (\boldsymbol{\mu}_i,\boldsymbol{\Omega}_i)$'s to be equal so that some data vectors share a common component $\mathrm{No}(\boldsymbol{\mu}_i, \boldsymbol{\Omega}_i)$ in the posterior predictive distribution. Thus, the posterior predictive distribution of the $\mathbf{X}_i$'s will tend to have fewer mixture components than data points, with the number of components being informed by the data. As $\alpha$ increases, the number of mixture components used to mimic the data generating distribution will also tend to increase. To allow the data to provide information about $\alpha$, we give $\alpha$ a Gamma prior as suggested by \citet{EscoWest95}. Because the Normal-Wishart base distribution $G_0$ is conjugate to the Gaussian distribution, the Collapsed Gibbs sampler proposed by \citet{Neal00} can be used to obtain posterior MCMC samples of the $\boldsymbol{\theta}_i$.

The prior elicitation for GDPM models can be challenging in general; the presence of high dimensional data greatly complicates this task. When $p \to \infty$ as $n$ remains fixed, \citet{ChanCana20} showed that the posterior predictive distribution tends to converge to either a single mixture component or $n$ distinct components, neither of which may approximate the true data distribution well. To enable accurate model fitting to datasets of arbitrary scales and distributions, we recommend the approach suggested by \citet{ShiMart19}, summarized in the following steps:
\begin{enumerate}
	\item Linearly transform the data vectors $\mathbf{X}_i$ to $\mathbf{Z}_i$ in a manner so that the $\mathbf{Z}_i$ have approximately zero mean and unit variance
	\item Apply the GDPM model to the $\mathbf{Z}_i$ with a specific choice of hyperparameters and obtain posterior inference on their data distribution
	\item Back-transform the inference on the $\mathbf{Z}_i$ to obtain inference on the original data $\mathbf{X}_i$
\end{enumerate}
For Step 2 of this procedure, we recommend using these values for the hyperparameters:
\begin{align*}
	\mathbf{m} = \mathbf{0}, 
	\quad v = 2p,
	\quad k =  \frac{v n \chi_p^2(\psi) (1-\psi)}{(v-p-1) (n-1) p}\\
	\quad \mathbf{W} = v \mathbf{I},
	\quad \psi = 0.99,
	\quad a = b = 1,
\end{align*}
where $\chi_p^2(\psi)$ is the $\psi^{th}$ quantile of the $\chi^2$ distribution with $p$ degrees of freedom. These values appear to work well when $n, p \le 250$ and were obtained using a combination of the method of moments approach in \citet{ShiMart19} and experimentation via simulations. \textcolor{black}{The development of a low information omnibus prior for Dirichlet process mixtures with high dimensional data is one of our working projects; we anticipate providing a systematic prior construction with mathematical properties in the near future.}

The empirical performance of the GDPM in modeling distributions was illustrated by \citet{ShiMart19}, where they used the GDPM to achieve accurate density estimation of both a mixture of bivariate t distributions and a Cauchy distribution, which does not have finite mean nor variance. Considering the capacity of GDPM, we anticipate good performance even with multimodal and heavy-tailed data-generating distributions.


\subsection{MCMC Sampling of the GDPM Model}
Because the distribution $G$ in the GDPM model is discrete, some $\boldsymbol{\theta}_i$'s are likely to be equal in the posterior distribution Thus, a clustering of the $n$ data vectors occurs such that their sampling distribution is modeled as a Gaussian mixture of $|\{\boldsymbol{\theta}_i: i=1, \dots, n\}|$ unique components. Let $\boldsymbol{c} = (c_1, \dots, c_n)$ denote the vector of labels for these unique mixture components among the data vectors. The process for updating $\boldsymbol{c}$ is achieved using the split-and-merge method introduced by \citet{JainNeal04}. In each iteration, we randomly pick two distinct observations, labeled $i$ and $l$. We represent the collection of other observations in the same cluster as either $i$ or $l$ as $\mathscr{C}$. In the scenario where $\mathscr{C}$ is found to be vacant, the simple split-merge technique is employed. Conversely, when $\mathscr{C}$ is not empty, we use the restricted Gibbs sampling split-merge. The split-merge algorithm involves a Metropolis-Hastings sampling phase. The probability of acceptance is given by: 
\begin{equation*}
	a(\boldsymbol{c}^{merge},\mathbf{c})=\min\left\{1, \frac{q(\boldsymbol{c}|\boldsymbol{c}^{merge})\mathrm{P}(\boldsymbol{c}^{merge})L(\boldsymbol{c}^{merge}|\mathbf{X})}{q(\boldsymbol{c}^{merge}|\boldsymbol{c})\boldsymbol{P}(\boldsymbol{c})L(\boldsymbol{c}|\mathbf{X})}
	\right\}\end{equation*} if $c_i \neq c_l$ where observation $i$ and $l$ are not in different clusters, and \begin{equation*}
	a(\mathbf{c}^{split}|\mathbf{c})=\min\left\{ 
	1, \frac{q(\boldsymbol{c}|\boldsymbol{c}^{split})\mathrm{P}(\mathbf{c}^{split})L(\mathbf{c}^{split}|\mathbf{X})}{q(\boldsymbol{c}^{split}|\boldsymbol{c})\mathrm{P}(\boldsymbol{c})L(\boldsymbol{c}|\mathbf{X})}
	\right\} \end{equation*} if $c_i = c_l,$ i.e., observation $i$ and $l$ are in the same cluster.

For the simple split-merge algorithm, $\frac{q(\mathbf{c}|\mathbf{c}^{merge})}{q(\mathbf{c}^{merge}|\mathbf{c})} =\frac{q(\mathbf{c}|\mathbf{c}^{split})}{ q(\mathbf{c}^{split}|\mathbf{c})} = 1$. For the restricted Gibbs sampling method, an initial launch state is generated randomly, which is then modified through multiple "intermediate" restricted Gibbs sampling steps to facilitate an effective division of the observations. The final launch state is then applied to calculate the transient probabilities. Further information on the split-and-merge procedure can be found in the \citet{JainNeal04}.

The ratio of priors, either $\mathrm{P}(\mathbf{c}^{merge})/\mathrm{P}(\mathbf{c})$ or $\mathrm{P}(\mathbf{c}^{split})/\mathrm{P}(\mathbf{c})$, depends on the partition function $\mathrm{P}(\mathbf{c})$. For Dirichlet process mixture with $K$ distinct clusters and concentration parameter $\alpha$, the distribution function of $\mathbf{c}$ can be expressed as $$
\mathrm{P}(\mathbf{c})=\alpha^K \frac{\prod_{c \in C}(n_c-1)!}{\prod_{j=1}^n(\alpha+j-1)},$$
where $C$ is the set of distinct cluster labels and $n_c$ is the number of observations with label $c.$ The formulas for calculating the likelihoods of splitting or merging clusters are presented as: 
\begin{align*}
	\frac{\mathrm{P}(\mathbf{c}^{split})}{\mathrm{P}(\mathbf{c})}&=\frac{\alpha(n_{c_1}-1)!(n_{c_2}-1)!}{(n_{c}-1)!};\\
	\frac{\mathrm{P}(\mathbf{c}^{merge})}{\mathrm{P}(\mathbf{c})}&=\frac{(n_{c}-1)!}{\alpha(n_{c_1}-1)!(n_{c_2}-1)!},\\
	\end{align*}
where $n_{c_1}$ and $n_{c_2}$ denote the counts of observations in the respective clusters.

The sampled values of cluster indices provide information on whether two observations are part of the same cluster, yet these values are not directly comparable across iterations. This lack of comparability is due to the "label switching" phenomenon, where identical index values might denote different clusters in separate iterations. In our work, we employ the method proposed by \citet{FritIcks09} for the summarization of posterior cluster labels derived from MCMC samples. We introduce a clustering estimate denoted as $c^*$ and calculate the probability that samples $i$ and $j$ are in the same cluster across $M$ MCMC samples with $\kappa_{ij}=\frac{1}{M}\sum_{m=1}^M \mathrm{I}(c_i^{(m)}=c_j^{(m)})$. To obtain a posterior cluster assignment, we aim to maximize the adjusted Rand index:

\begin{align*}
	\mathrm{AR}&(c^*,\boldsymbol{\kappa}) = \\
&\frac{\sum_{i<j}\mathrm{I}_{\{c_i^*=c_j^*\}}\kappa_{ij}-\sum_{i<j}\mathrm{I}_{\{c_i^*=c_j^*\}}\sum_{i<j}\kappa_{ij}/C_2^N}
	{\frac{1}{2}[\sum_{i<j}\mathrm{I}_{\{c_i^*=c_j^*\}}+\sum_{i<j}\kappa_{ij}]-\sum_{i<j}\mathrm{I}_{\{c_i^*=c_j^*\}}\sum_{i<j}\kappa_{ij}/C_2^N}.
\end{align*}

This approach effectively addresses the label-switching problem and is straightforward to implement using the R package "mcclust" \citep{Frit12}.

\subsection{Knockoff Generation}

By implementing the method described by \citet{Frit12}, we identify $K$ unique clusters. Sampling a new parameter set $\boldsymbol{\theta}=(\boldsymbol{\mu},\boldsymbol{\Omega})$ involves drawing from a specific categorical distribution expressed as  $$\boldsymbol{\theta} \sim \sum_{k=1}^K \frac{n_k}{n+\alpha} \delta(\boldsymbol{\theta}_k)+\frac{\alpha}{n+\alpha}\boldsymbol{\theta}_0,$$
where $n_k$ represents the count of observations within the $k$th cluster, $\boldsymbol{\theta}_k$ denotes parameters sampled from the posterior distribution of observations attributed to cluster $k$, and $\boldsymbol{\theta}_0$ originates from the base Normal-Wishart distribution. Consequently, the data-generating distribution for the $i$th observation is given by
$$f(\mathbf{x}_i)= \sum_{k=1}^K \frac{n_k}{n+\alpha} f(\mathbf{x}_i|\boldsymbol{\theta}_k)+\frac{\alpha}{n+\alpha} f(\mathbf{x}_i|\boldsymbol{\theta}_0).
$$

To generate a knockoff for the $j$th variable of the $i$th row, we follow the formula:
	\begin{align*}
		f(\widetilde{\mathbf{x}}_{ij}|\mathbf{x}_{i,-j})\propto& \sum_{k=1}^K \left[\frac{n_k}{n+\alpha} 
		f(\mathbf{x}_{i,-j}|\boldsymbol{\theta}_k)\right]f(\widetilde{\mathbf{x}}_{ij}|\mathbf{x}_{i,-j},\boldsymbol{\theta}_k)\\
		&+\left[\frac{\alpha}{n+\alpha}f(\mathbf{x}_{i,-j}|\boldsymbol{\theta}_0)\right]
		f(\widetilde{\mathbf{x}}_{ij}|\mathbf{x}_{i,-j},\boldsymbol{\theta}_0).\\
	\end{align*}
In this context, for all $k$ ranging from $0$ to $K$, $f(\mathbf{x}_{i,-j}|\boldsymbol{\theta}_k)$ is the marginal distribution for the remaining $p-1$ variables and \\ $f(\widetilde{\mathbf{x}}_{ij}|\mathbf{x}_{i,-j},\boldsymbol{\theta}_k)$ specifies the conditional distribution. This mechanism for knockoff generation is outlined in Algorithm 1.

\begin{algorithm}
	\caption{Steps for sampling GDPM knockoff}
	\begin{algorithmic}[1] 
		\FOR{$i=1$ to $N$}
		\FOR{$j=1$ to $p$}
		\FOR{$k=0$ to $K$}
		\STATE{Sample $\boldsymbol{\theta}_k$, either from the base distribution $(k=0)$ or the posterior distribution with observations assigned to cluster $k$;}
		\STATE {Compute the multivariate Gaussian $f(\mathbf{x}_{i,-j}|\boldsymbol{\theta}_k)$ likelihood for the rest $p-1$ columns;}
		\STATE {Sample $x_{ijk}\sim f(\cdot|\mathbf{x}_{i,-j},\boldsymbol{\theta}_k)$ from the conditional Gaussian distribution.}
		\ENDFOR
		\STATE{Sample $x_{ij}\sim \frac{1}{B}\sum_{k=0}^K w_{ijk}\delta(x_{ijk}).$ The weights $w_{ijk}$ is either $\frac{n_k}{n+\alpha}f(\mathbf{x}_{i,-j}|\boldsymbol{\theta}_k)$ or $\frac{\alpha}{n+\alpha}f(\mathbf{x}_{i,-j}|\boldsymbol{\theta}_0),$ and $B$ is the normalizing constant.}
		\ENDFOR
		\ENDFOR
	\end{algorithmic}
\end{algorithm}
Any choice of importance statistic may be applied to knockoffs generated from the GDPM model, provided that the statistic satisfies the antisymmetry property \citep{CandFan18}. The most common choices involve a lasso model and use either the lasso signed max (LSM) statistic, which is the $\pm 1$ indicator on whether the feature enters the model before its knockoff copy multiplied by the largest penalty parameter at which the feature or its knockoff counterpart enters the model, or the lasso coefficient difference (LCD), which measures the difference in magnitude of the coefficient estimates for the original and knockoff features.

\section{ROBUSTNESS OF GDPM KNOCKOFF GENERATOR IN LARGE SAMPLES}
\citet{CandFan18} showed that the respective knockoff and knockoff+ filtering procedures will control the modified FDR and ordinary FDR at a prespecified level $q$, provided two conditions hold: (1) the knockoffs are constructed independently of the outcome, i.e. $\widetilde{\mathbf{X}} \perp Y | \mathbf{X}$; and (2) a pairwise exchangeability condition holds such that, for any subset $\mathcal{A} \subset \{1,\dots,p\},$ \textcolor{black}{after swapping the original features in $\mathcal{A}$ with their knockoffs,} $(\mathbf{X},\widetilde{\mathbf{X}})_{\text{swap}(\mathcal{A})}$ and $(\mathbf{X}, \widetilde{\mathbf{X}})$ have the same distribution. In the following text, we will use $\widehat{S}$ and $H_0$ to denote the selected feature set and the set of null features, respectively. Vertical bars are used to indicate the cardinality of the set.

Condition (1) is automatically satisfied if the knockoffs are created without using any outcome data. \textcolor{black}{Given other predictors $X_{-j},$ when the conditional distribution for each original feature $X_j$ and its knockoff $\widetilde{X}_j, j=1,\dots,p,$ are identical, i.e. the knockoff conditional model matches the true conditional distribution exactly, Condition (2) holds. But in practice, the knockoff conditional is an approximation estimated from existing data. Therefore, FDR control may not be strictly guaranteed under such approximations.}

To investigate to what degree the FDR may be inflated under knockoff generators that approximate the true conditional distribution, \citet{BarbCand20} considered the setting where an unlabeled dataset of $m$ observations that are independent from $\{(\mathbf{X}_i,Y_i)\}_{i=1}^n$ is used to construct the knockoff generating model. Their work obtained a bound on the FDR and showed that the degree to which this bound exceeds the targeted FDR level of $q$ depends on how much the approximated conditional distribution deviates from the true feature conditional distribution. This endorses the use of our GDPM model, because GDPM models have been shown to provide posterior predictive density functions that consistently estimate the true data density for various classes of distributions under certain regularity conditions \citep{Tokd06,WuGhos08,WuGhos10}. Based on the findings of \citet{BarbCand20} and \citet{WuGhos10}, we provide a large sample guarantee of FDR control for the GDPM knockoffs under one class of true feature distributions. 

Assume there are $n$ observations and $p$ features, and the feature distribution is continuous. We wish to generate knockoffs from a distribution with the same support. \citet{BarbCand20} investigated the robustness of FDR control under knockoff filtering when the model used to generate knockoffs is misspecified. In their work, they let $p_X^*$ and $p_j^*$ denote the true joint density of the original features $\mathbf{X}$ and the conditional densities of each $X_j$, $j=1,2,\dots, p$. They assume that these densities are estimated by a joint density $p_X$ and some conditional densities $p_j$ for the features; these densities are estimated independently from the labeled dataset $\{(\mathbf{X}_i,Y_i)\}_{i=1}^n$, for instance, using a separate unlabeled dataset or expert opinion. The difference in the estimated and true conditional densities for feature $j$ is measured using a surrogate for the Kullback-Leibler divergence that is defined as follows:
\begin{align*}
	\widehat{KL}_j = \sum_{i=1}^n \log\left( \frac{p_j^*(\mathbf{X}_{ij} | \mathbf{X}_{i, -j}) \cdot p_j(\widetilde{\mathbf{X}}_{ij} | \mathbf{X}_{i, -j})}{p_j(\mathbf{X}_{ij} | \mathbf{X}_{i, -j}) \cdot p_j^*(\widetilde{\mathbf{X}}_{ij} | \mathbf{X}_{i, -j})}\right).
\end{align*}
They showed that controlling the size of this divergence for null features will provide a bound on the FDR under knockoff filtering, as stated in their Theorem 1.

\begin{thm1}
	Consider the null variables for which $\widehat{KL}_j < \epsilon$. Using the knockoff+ filter with a targeted FDR rate of $q$, the FDR is bounded as
	\begin{align*}
		FDR& = E\left(\frac{|\widehat{S} \cap H_0|}{|\widehat{S}| \vee 1} \right)\\
		&\le \min_{\epsilon \ge 0} \left\{q e^\epsilon + P\left(\max_{j \in H_0} \widehat{KL}_j > \epsilon \right) \right\}.
	\end{align*}
	
	Similarly, using the knockoff filter with a targeted modified FDR rate of $q$, the modified FDR is bounded as
	\begin{align*}
		mFDR &= E\left(\frac{|\widehat{S} \cap H_0|}{|\widehat{S}| + q^{-1}} \right)\\
		&\le \min_{\epsilon \ge 0} \left\{ q e^\epsilon + P\left(\max_{j \in H_0} \widehat{KL}_j > \epsilon \right) \right\}.
	\end{align*}
	
\end{thm1}

We proceed to show a large sample guarantee of FDR control when knockoffs are generated from a GDPM model. Assume that $m$ feature vectors are used to fit this model and that these vectors have the same distribution as $\mathbf{X}$. By Theorem 1 of \citet{BarbCand20}, it suffices to show that for any given $\epsilon$, there exists $M_\epsilon$ such that $P(\widehat{KL}_j > \epsilon) = 0$ when $m \ge M_\epsilon$. We show this using results from \citet{WuGhos10}, who showed $L_1$-consistency of GDPM models for the true data distribution under certain regularity conditions.

\begin{thm2}
	Suppose that conditions [B5-B7] of \citet{WuGhos10} are satisfied. Then there exists $M_\epsilon$ such that $
	FDR \le q + \epsilon$ when $m \ge M_\epsilon$. Consequently, the FDR is guaranteed to be bounded arbitrarily close to $q$ under the Knockoff+ filter, provided that $m$ is sufficiently large. Similarly, the modified FDR can be bounded arbitrarily close to q under the Knockoff filter, provided that m is large enough. 
	\label{theorem1}
\end{thm2}

The details of Theorem 1 are presented in the appendix.

\section{SIMULATION STUDY}
\label{sec:sims}

In this section, we conduct a comprehensive comparison between our nonparametric Bayesian GDPM knockoff generator, the second-order Gaussian knockoff generator, and the Gaussian finite mixture knockoff generator \citep{GimeGhor19}, focusing on their performance in controlling the FDR and enhancing statistical power. For each method, we derive empirical estimates of both FDR and power by evaluating the counts of null and non-null features identified after employing the knockoffs and the more conservative knockoffs+ filters for achieving approximate and precise FDR control, respectively. The knockoff+ refers to \citet{CandFan18}'s selection with offset, $$
\widehat{\mathrm{FDR}}_{+}(\lambda)=\frac{1+{|\widetilde{S}(\lambda)|}}{|\widehat{S}(\lambda)|\vee 1},$$ where $\widetilde{S}(\lambda)$ is the number of knockoff features selected and $\widehat{S}(\lambda)$ is the number of original features selected.  The target FDR was set to be $0.2,$ $0.1$ and $0.05$ respectively.


\begin{table*}[t!]
	\begin{tabular}{p{0.1cm}p{0.1cm}c}
		\multicolumn{3}{c}{\small
			\hspace{1.5cm} \parbox{0.2\textwidth}{One cluster covariance matrix $\mathbf{I}$}
			\hspace{0.5cm} \parbox{0.2\textwidth}{One cluster\\ covariance matrix $\mathrm{AR}(0.5)$}
			\hspace{0.5cm} \parbox{0.2\textwidth}{Mixture of two clusters\\ covariance matrix $\mathbf{I}$}
			\hspace{0.5cm} \parbox{0.2\textwidth}{Mixture of two clusters\\ covariance matrix $\mathrm{AR}(0.5)$}}\\
		\multirow{2}{*}{\rotatebox[origin=c]{90}{\small Target FDR=0.2}} & 
		\rotatebox[origin=c]{90}{\hspace{2cm} \small FDR }&
		\includegraphics[width=0.23\textwidth]{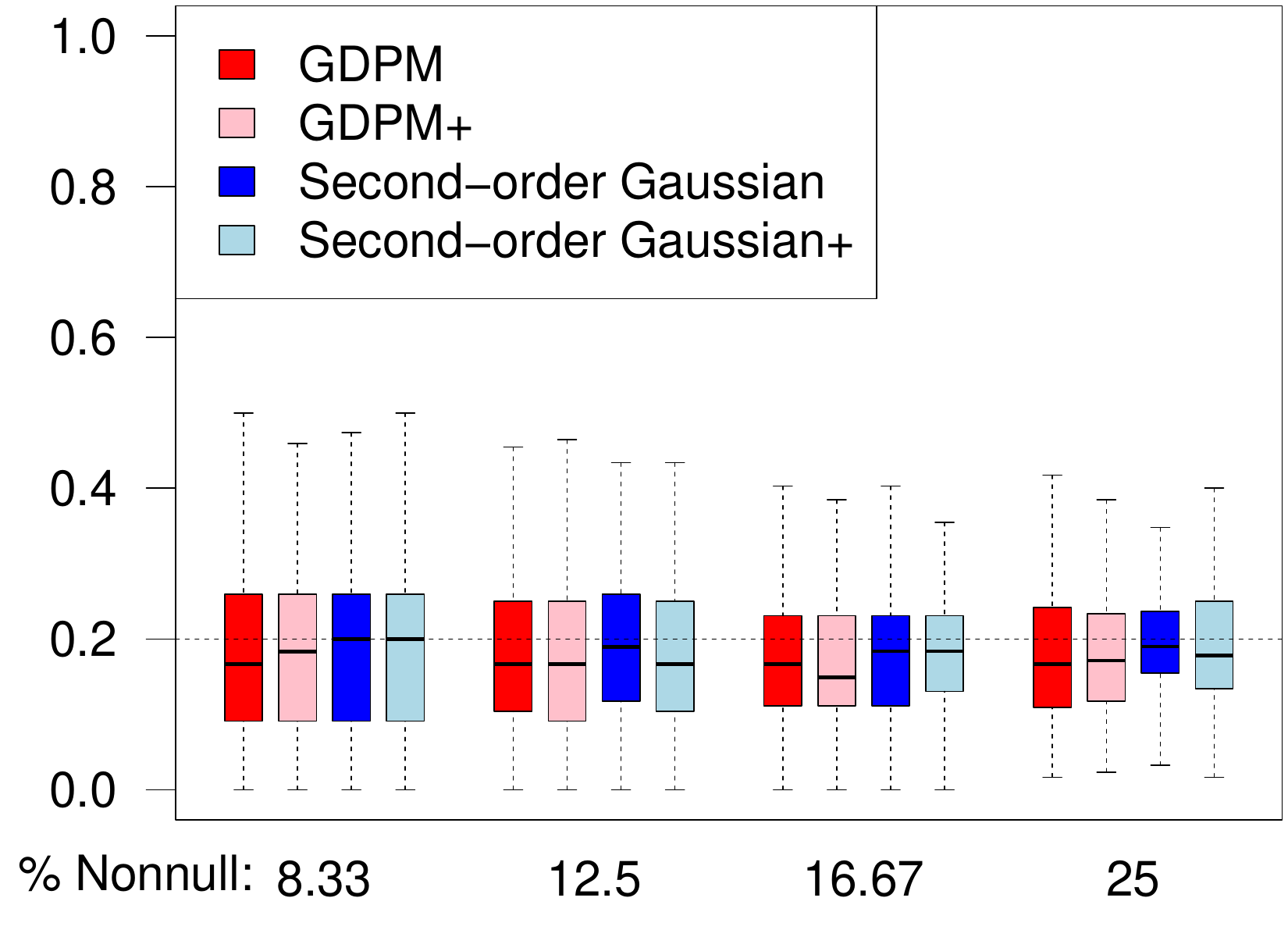}
		\includegraphics[width=0.23\textwidth]{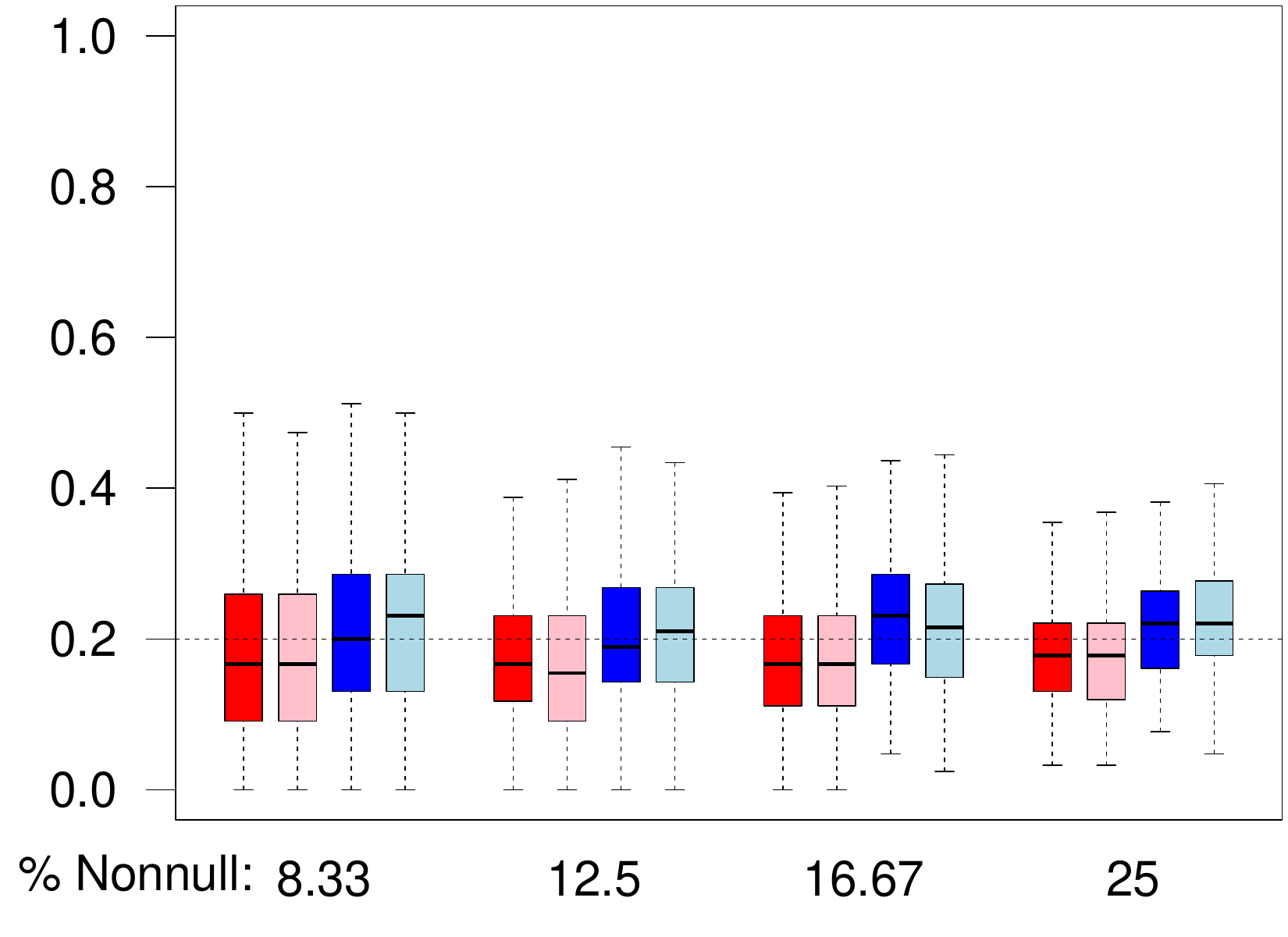}
		\includegraphics[width=0.23\textwidth]{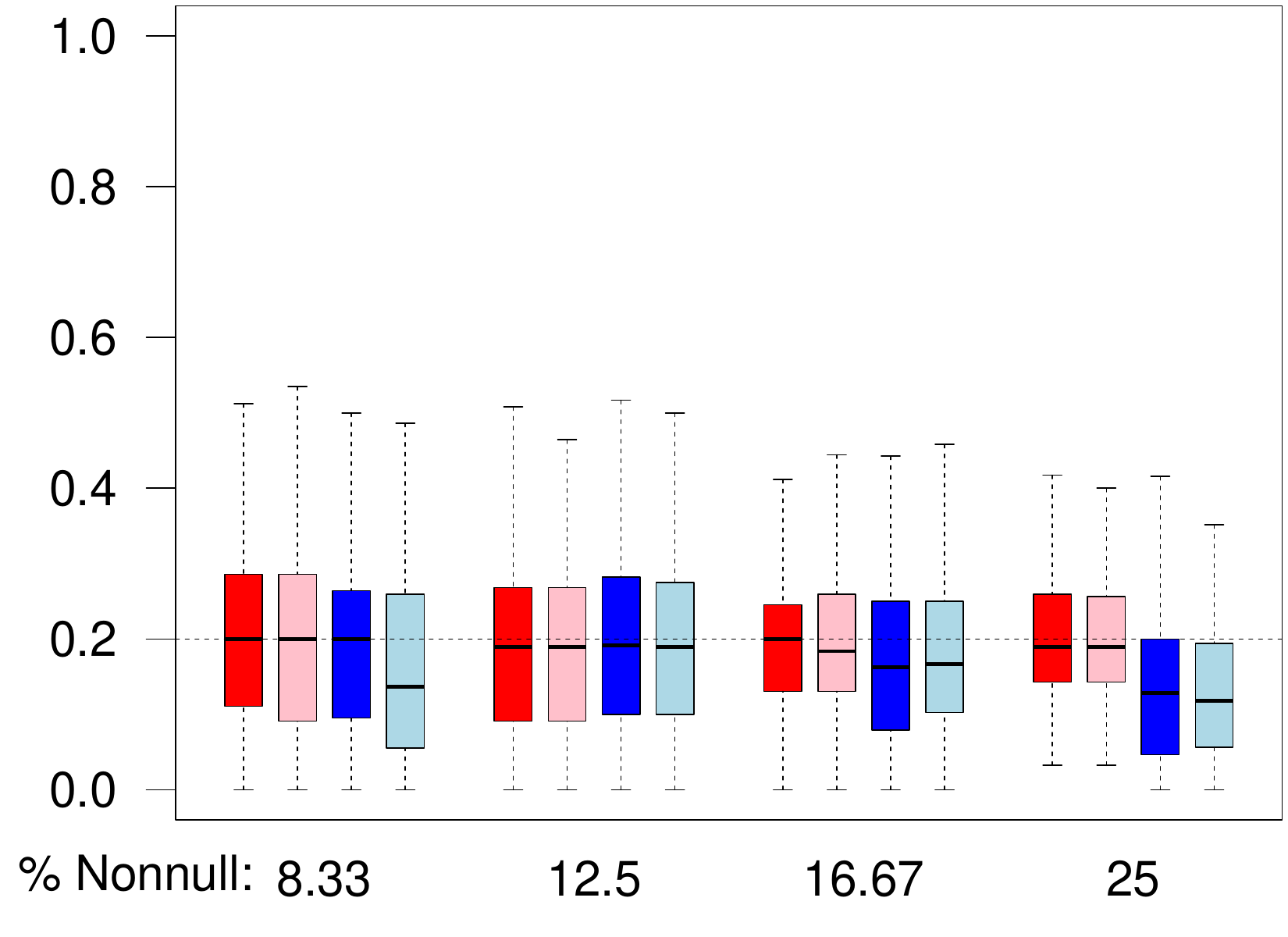}
		\includegraphics[width=0.23\textwidth]{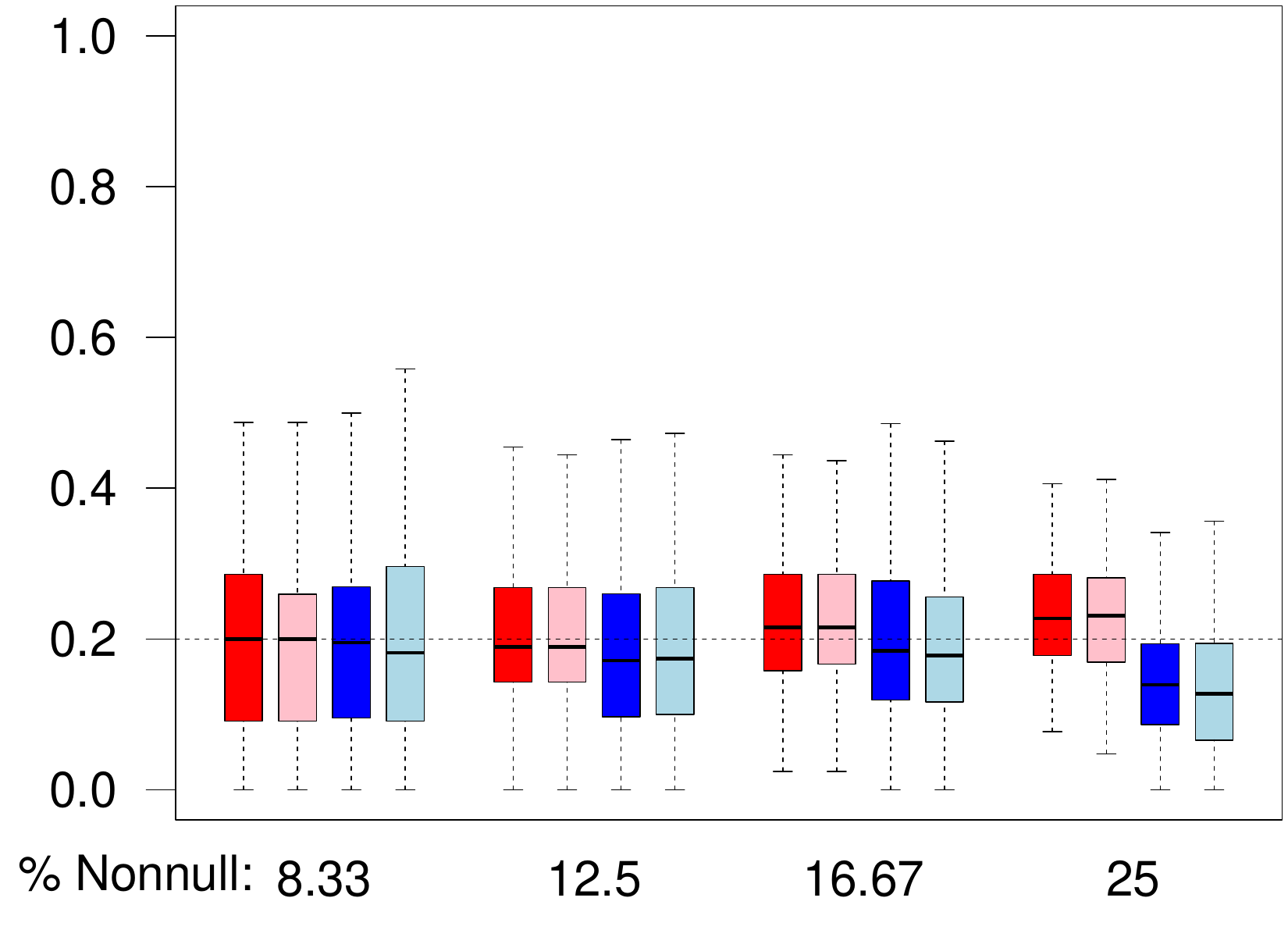}\\[-1cm]
		&\rotatebox[origin=c]{90}{\hspace{2cm}\small Power}&
		\includegraphics[width=0.23\textwidth]{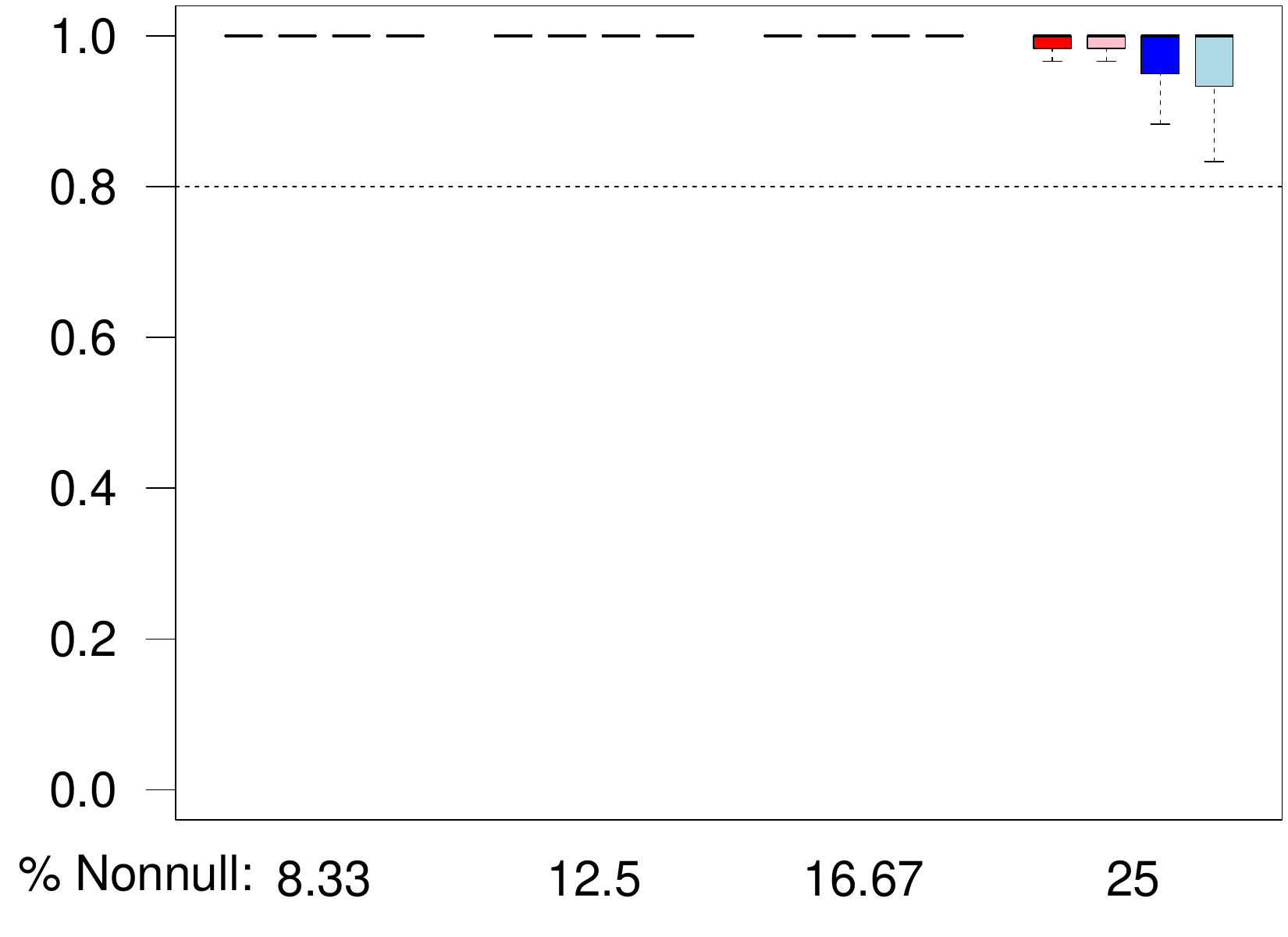}
		\includegraphics[width=0.23\textwidth]{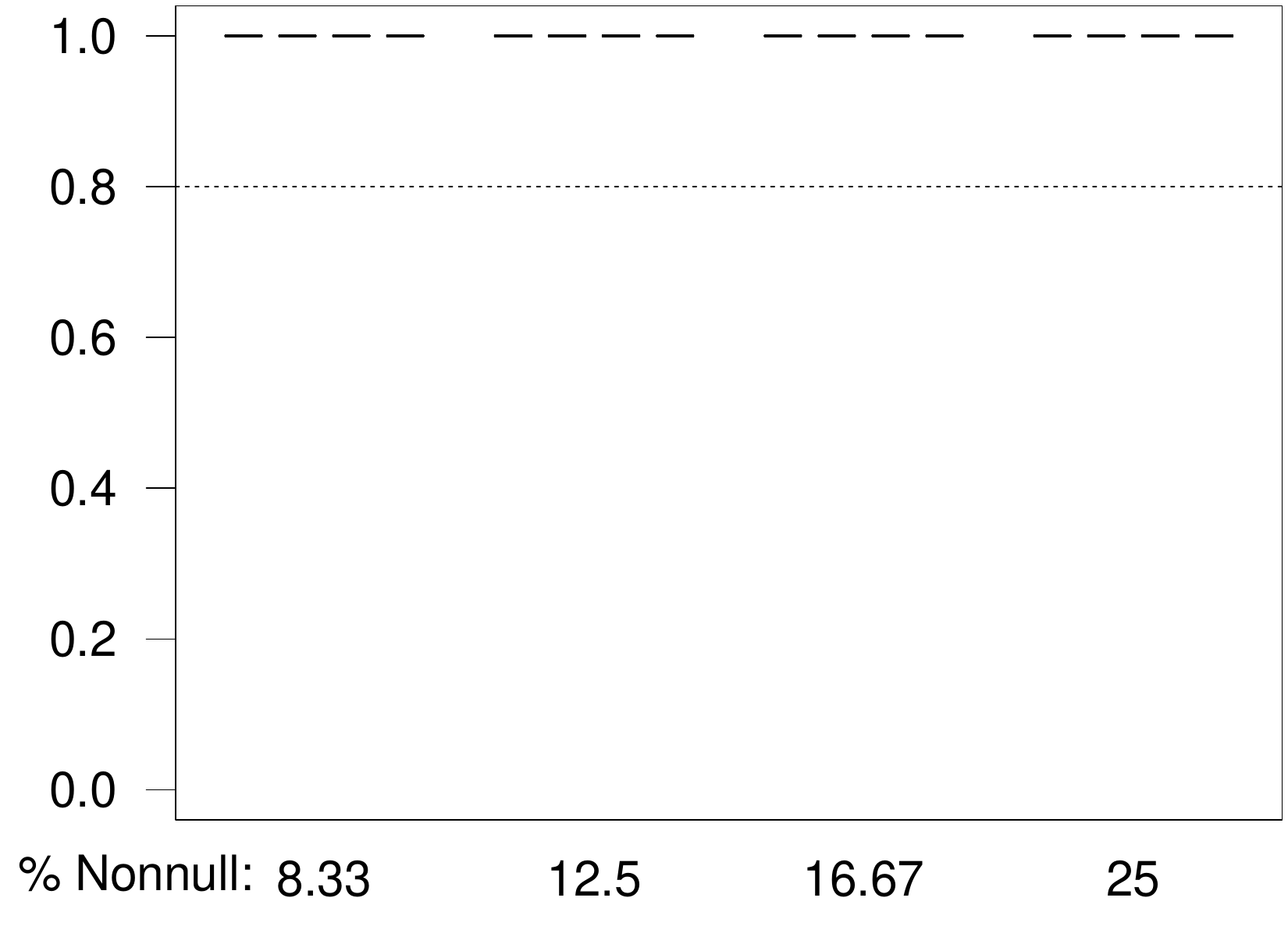}
		\includegraphics[width=0.23\textwidth]{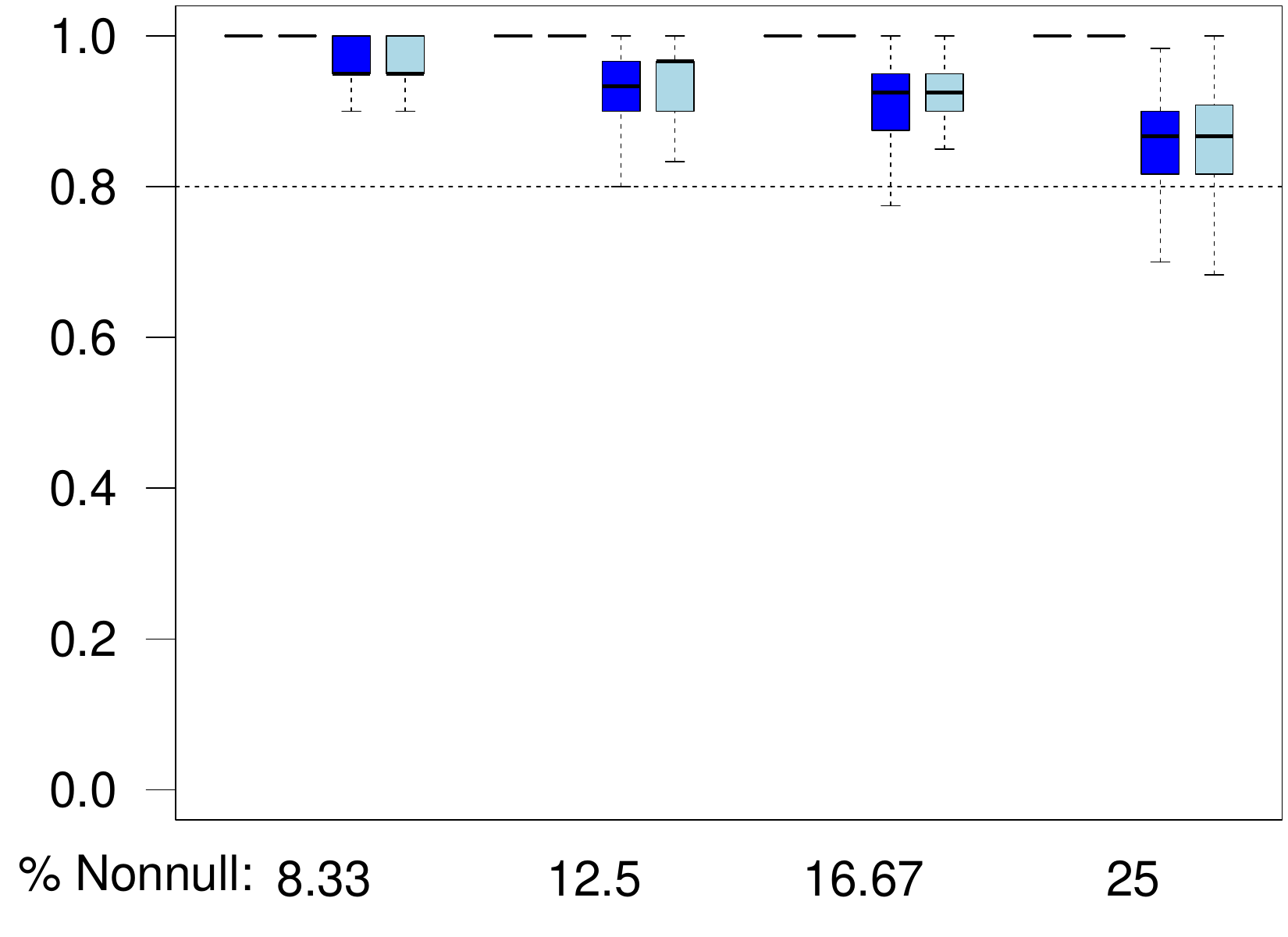}
		\includegraphics[width=0.23\textwidth]{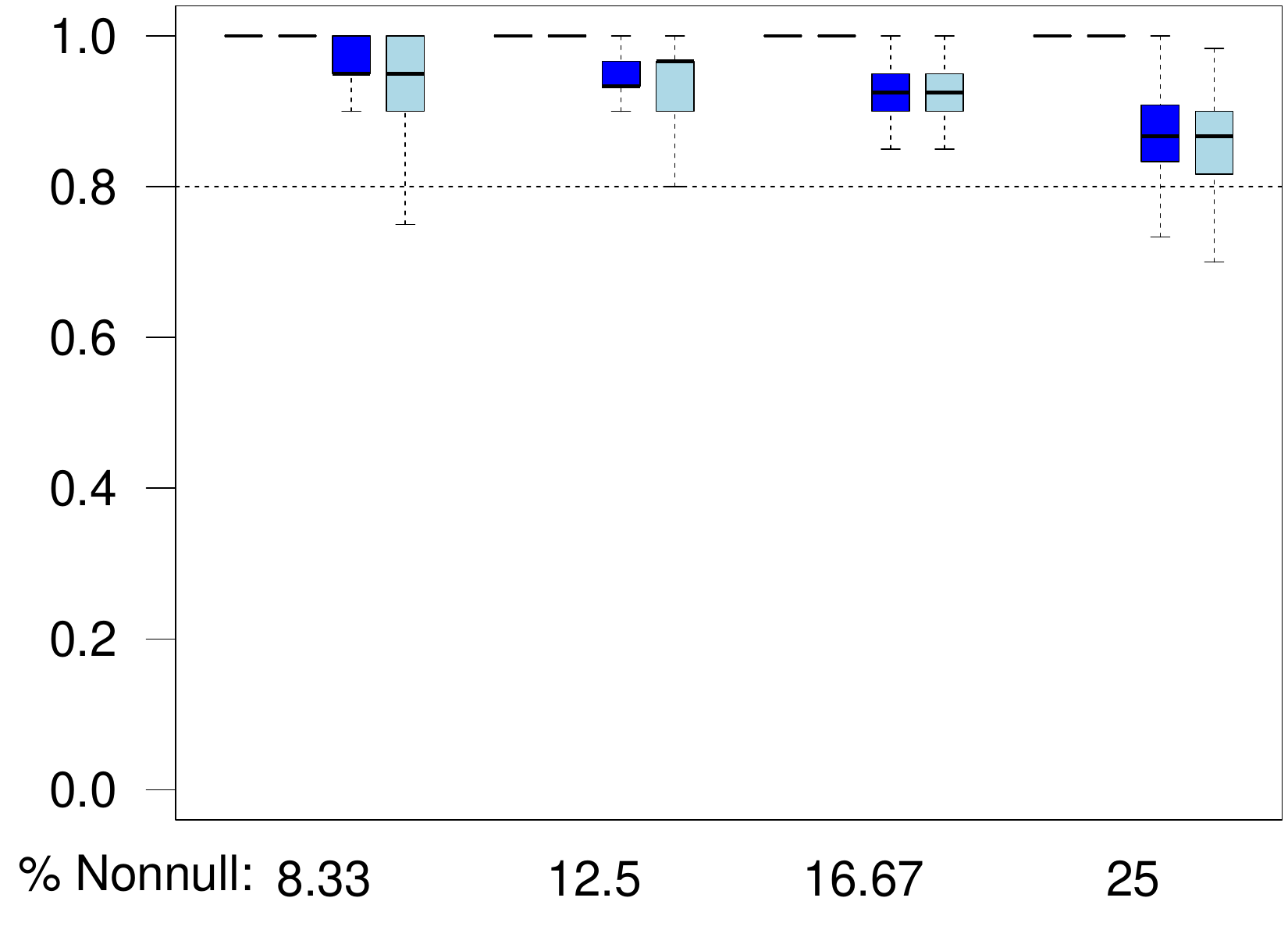}\\[-1cm]
	\end{tabular}
	\begin{tabular}{p{0.1cm}p{0.1cm}c}
		\multirow{2}{*}{\rotatebox[origin=c]{90}{\small Target FDR=0.1}} & 
		\rotatebox[origin=c]{90}{\hspace{2cm}\small FDR }&
		\includegraphics[width=0.23\textwidth]{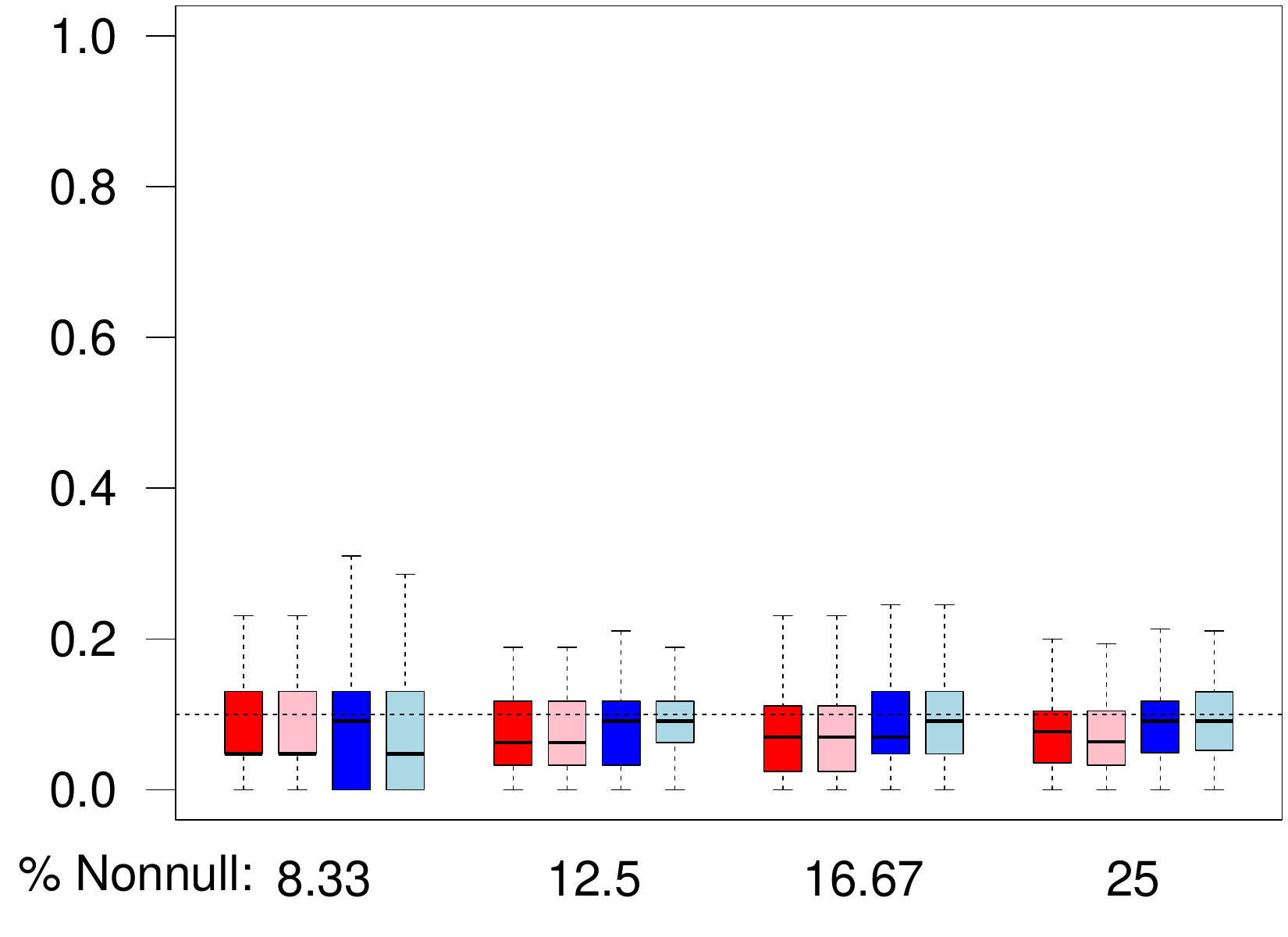}
		\includegraphics[width=0.23\textwidth]{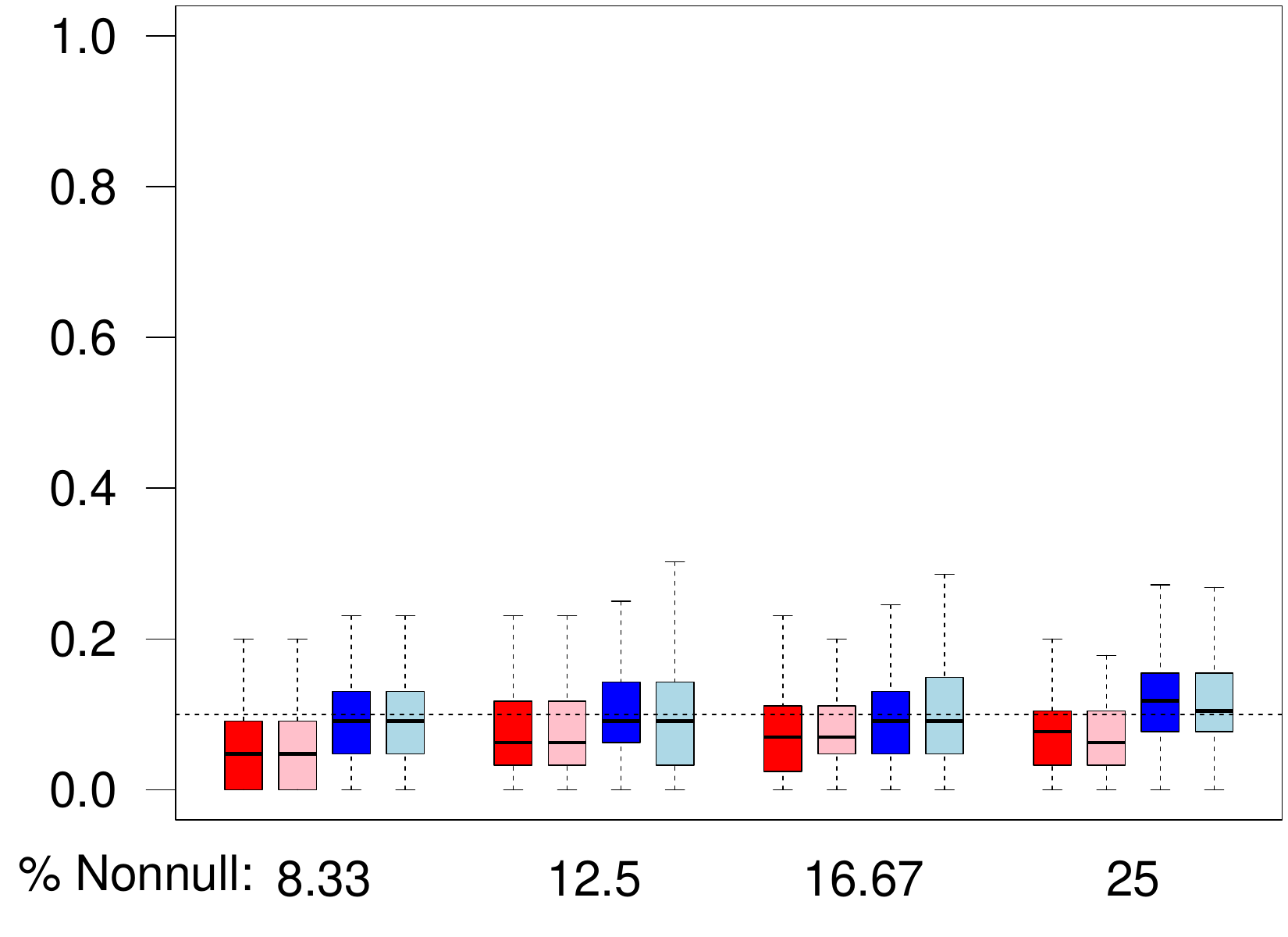}
		\includegraphics[width=0.23\textwidth]{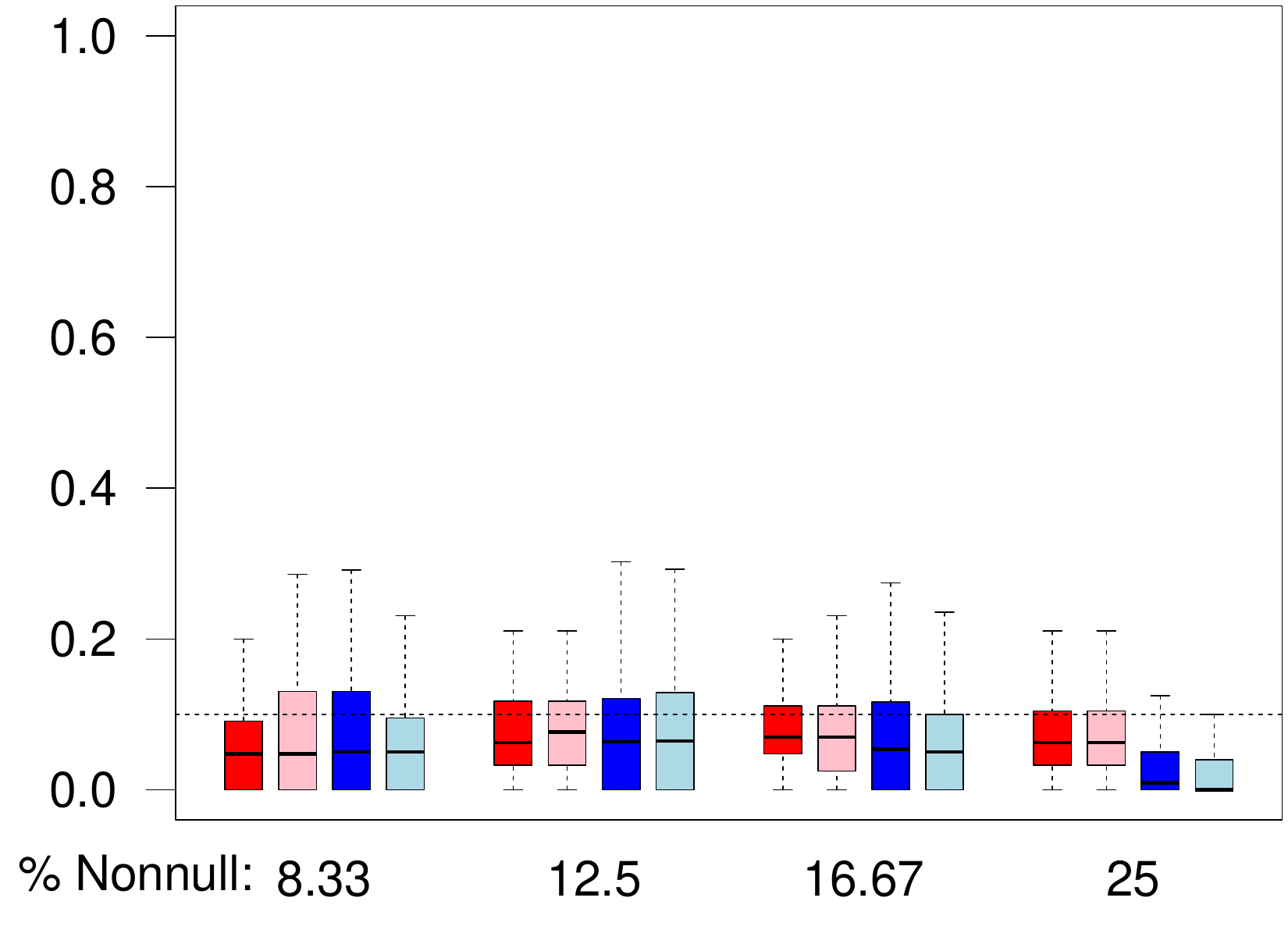}
		\includegraphics[width=0.23\textwidth]{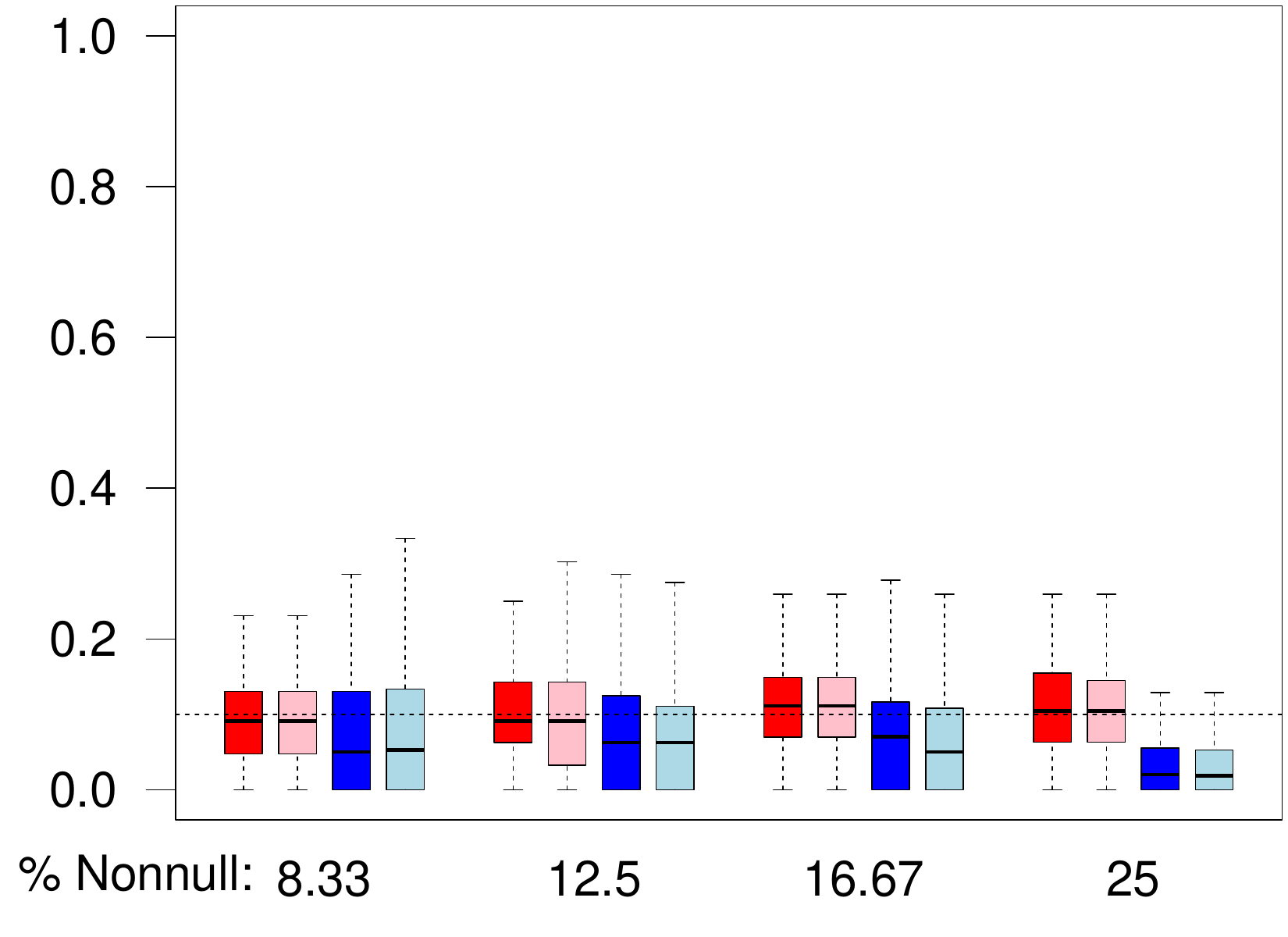}\\[-1cm]
		&\rotatebox[origin=c]{90}{\hspace{2cm}\small Power}&
		\includegraphics[width=0.23\textwidth]{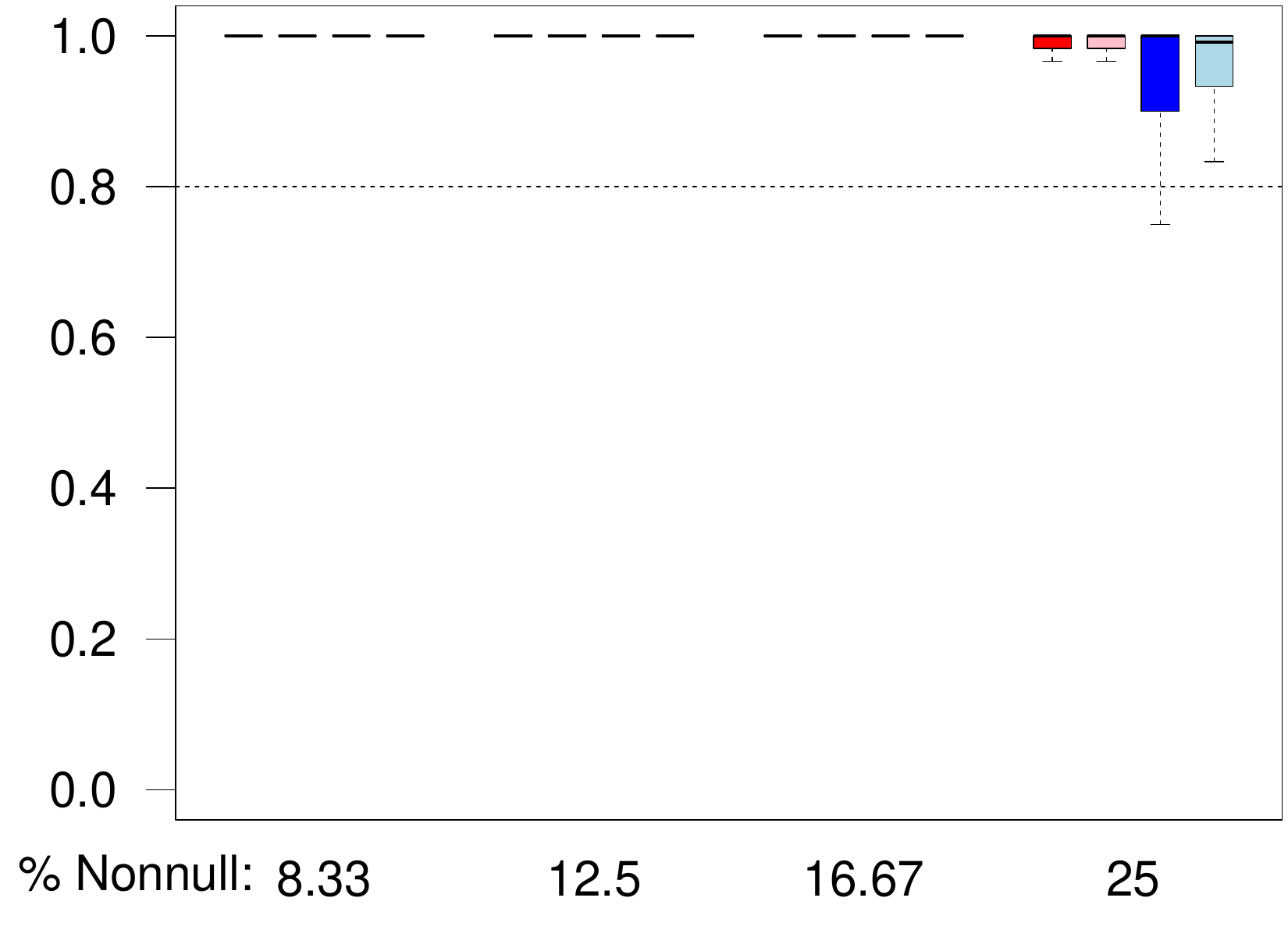}
		\includegraphics[width=0.23\textwidth]{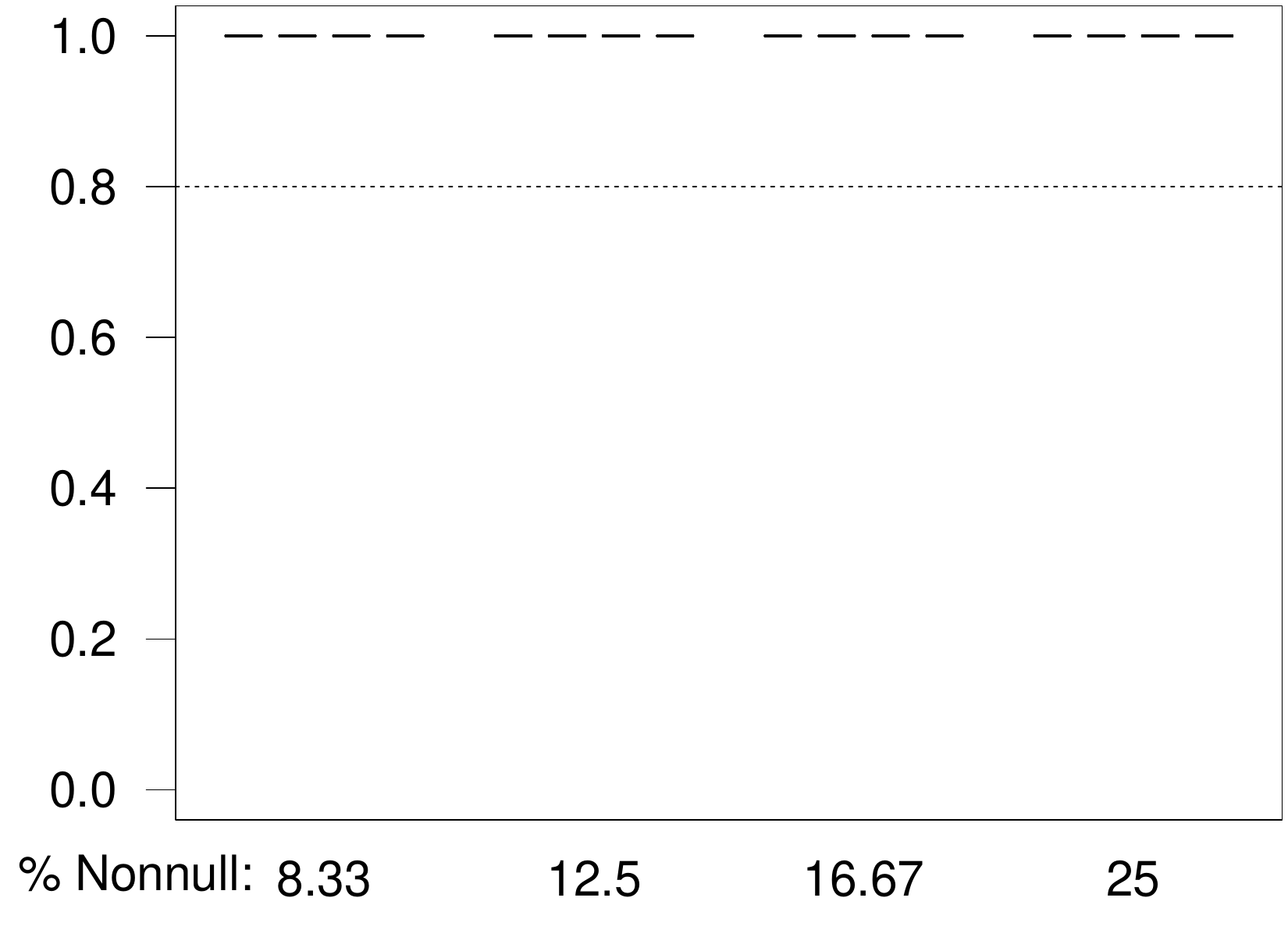}
		\includegraphics[width=0.23\textwidth]{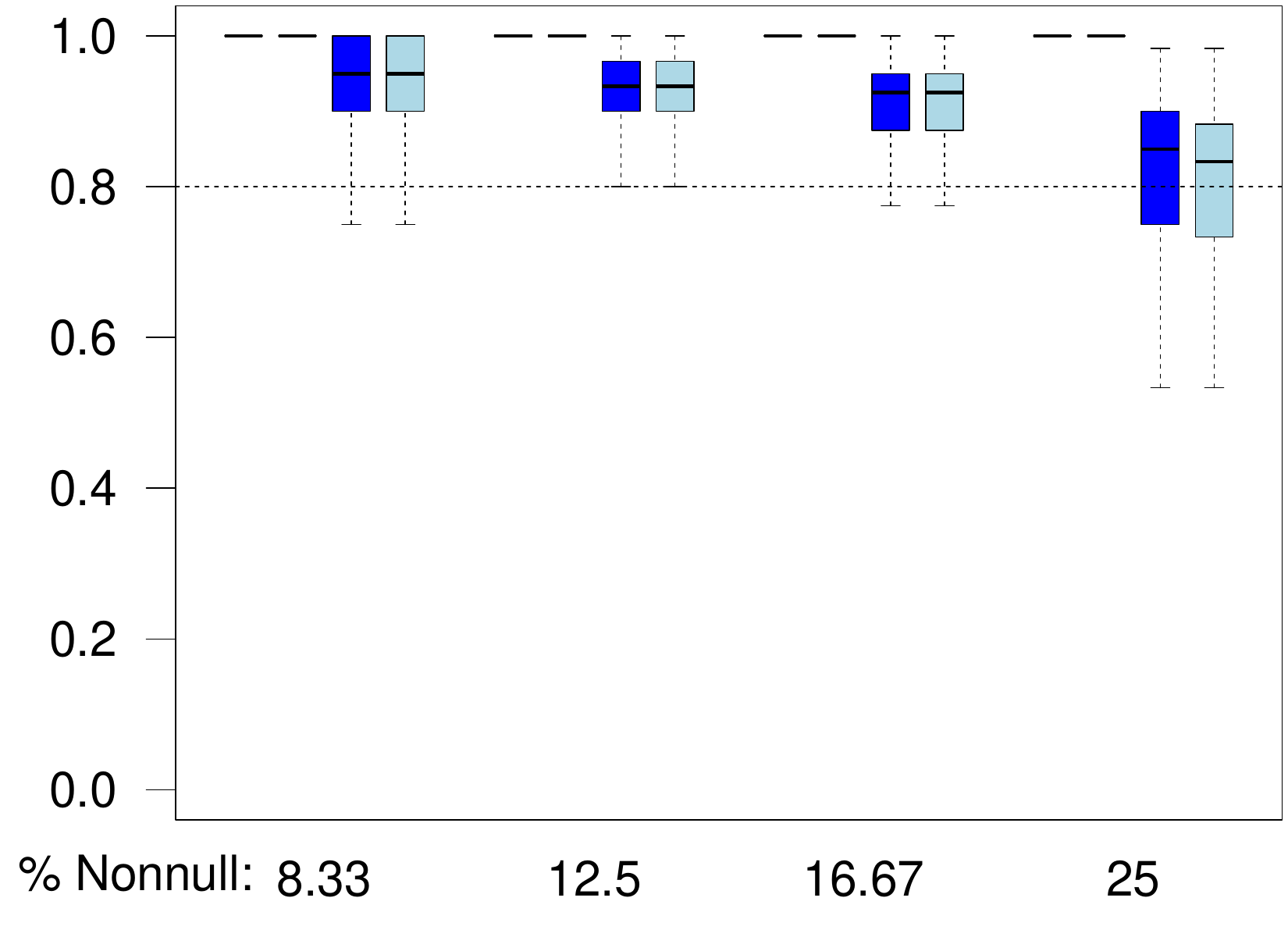}
		\includegraphics[width=0.23\textwidth]{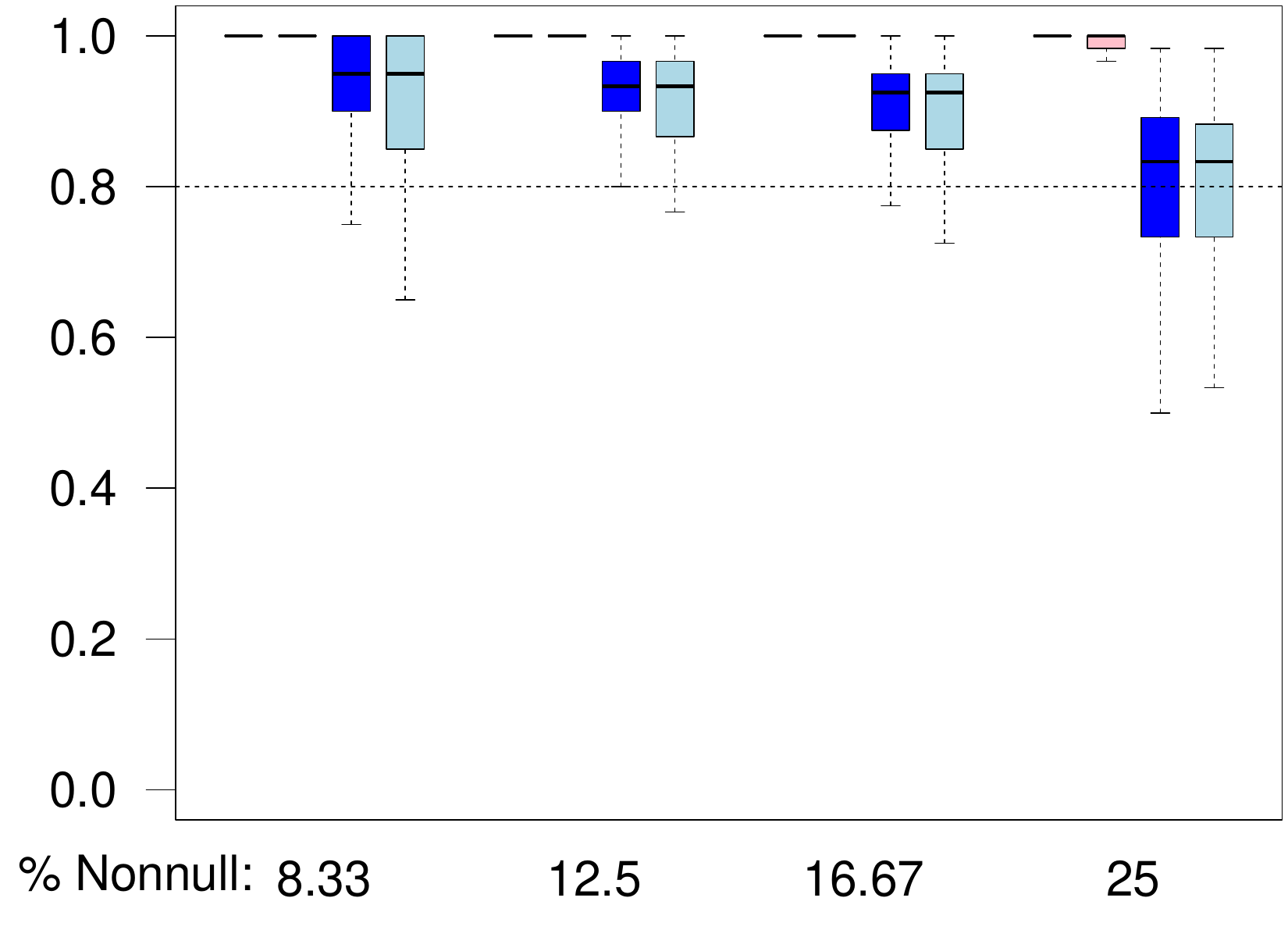}\\[-1cm]
	\end{tabular}
	\begin{tabular}{p{0.1cm}p{0.1cm}c}
		\multirow{2}{*}{\rotatebox[origin=c]{90}{\small Target FDR=0.05}} & 
		\rotatebox[origin=c]{90}{\hspace{2cm}\small FDR} &
		\includegraphics[width=0.23\textwidth]{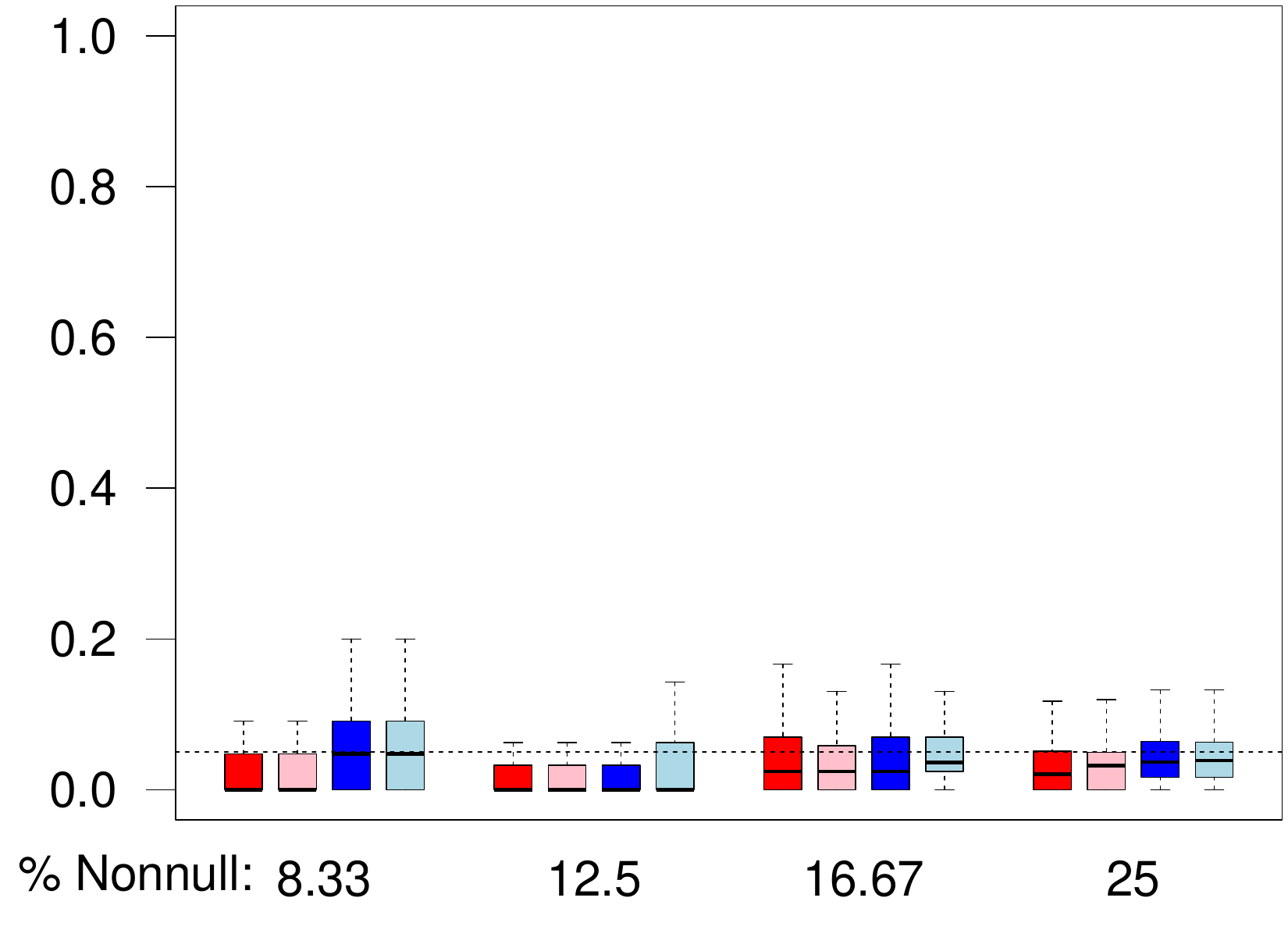}
		\includegraphics[width=0.23\textwidth]{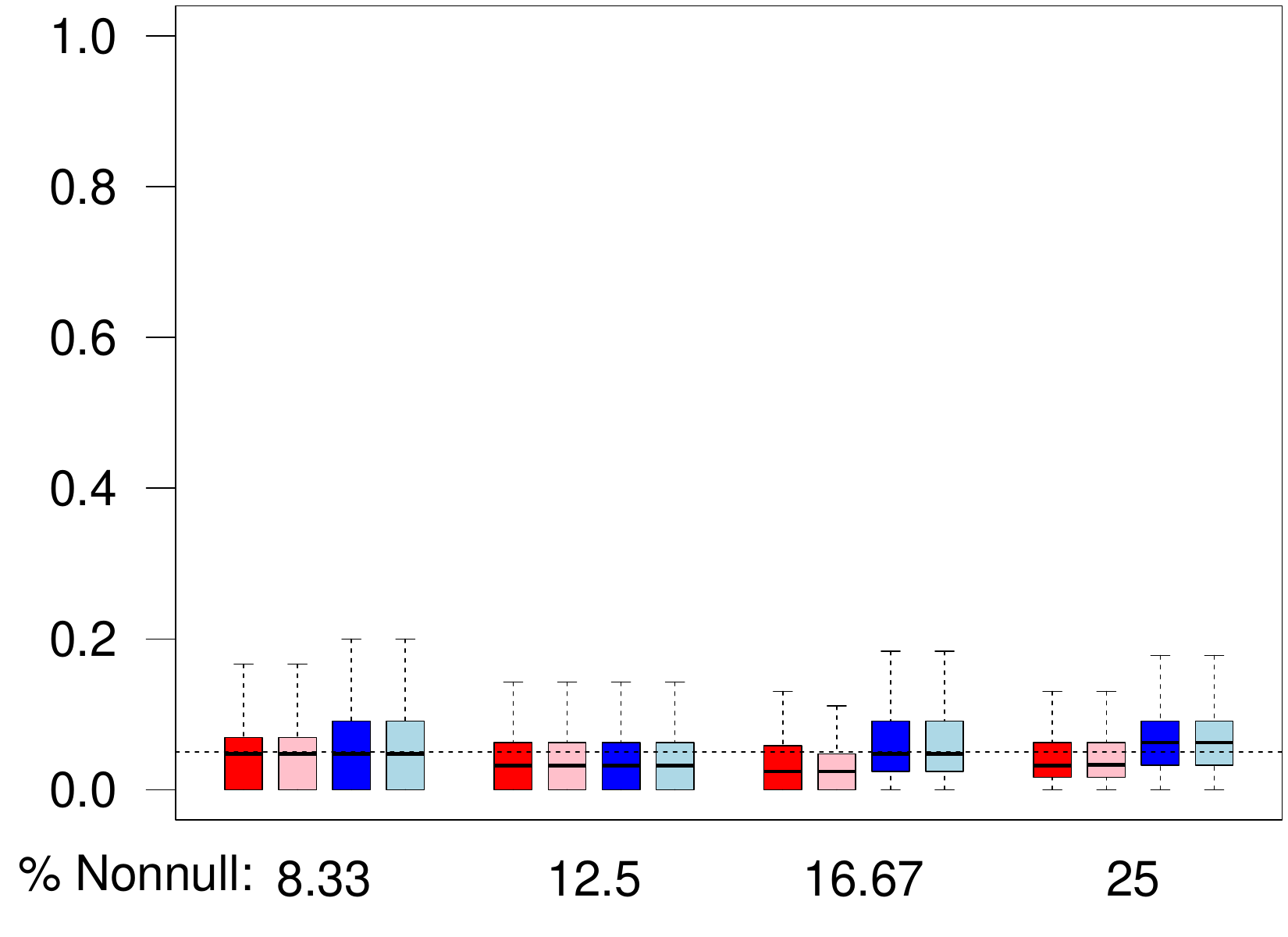}
		\includegraphics[width=0.23\textwidth]{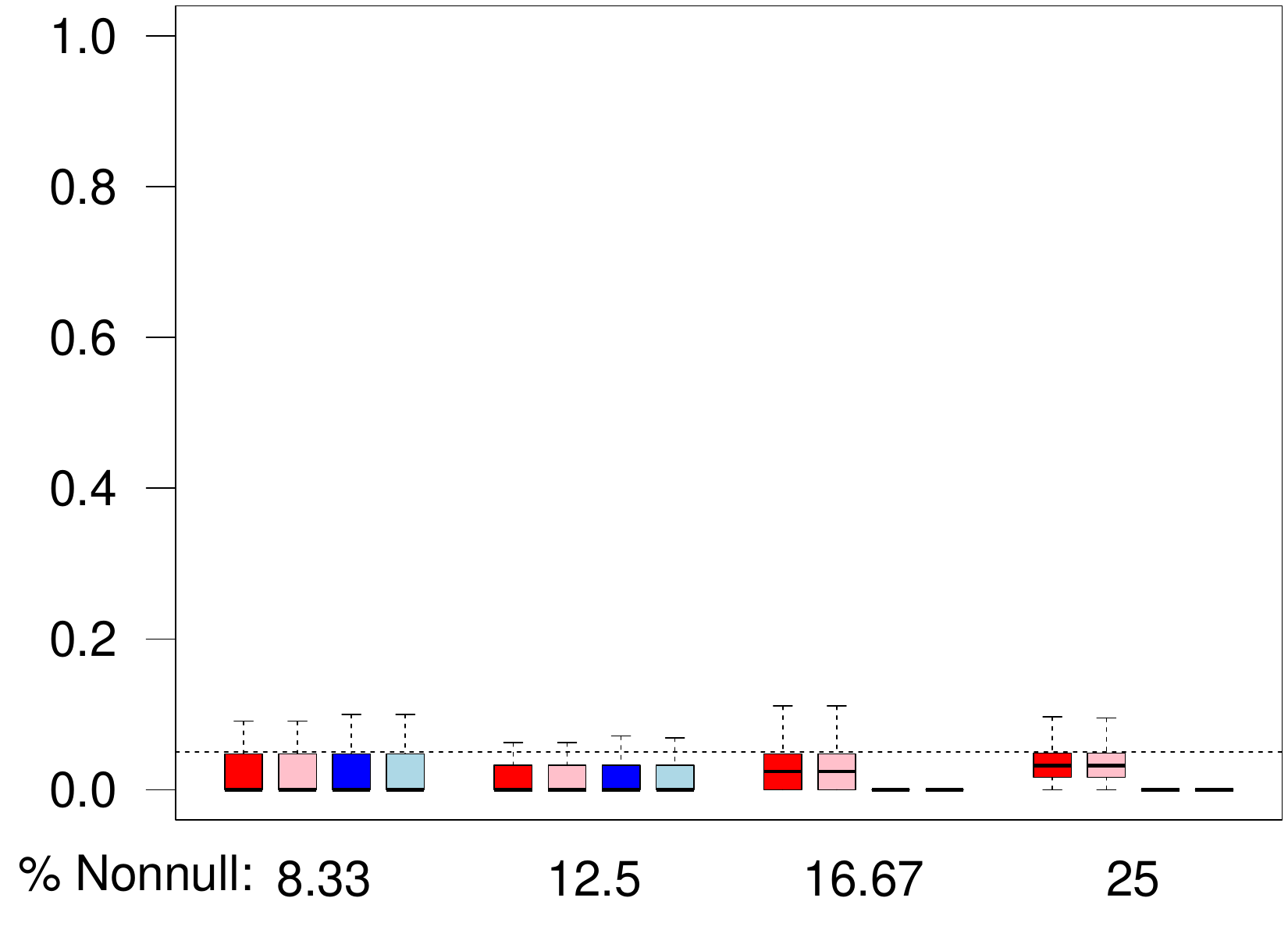}
		\includegraphics[width=0.23\textwidth]{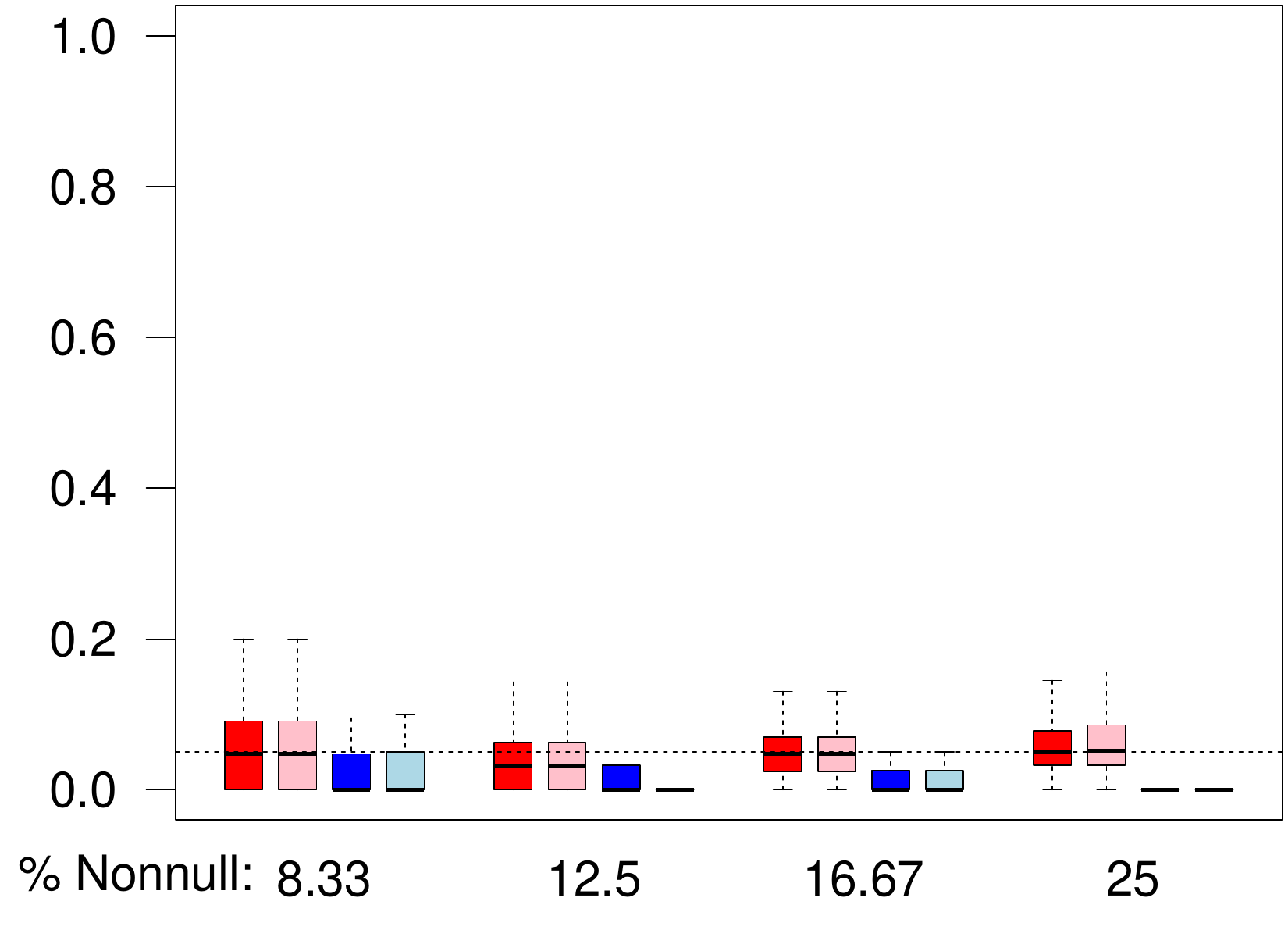}\\[-1cm]
		&\rotatebox[origin=c]{90}{\hspace{2cm}\small Power}&
		\includegraphics[width=0.23\textwidth]{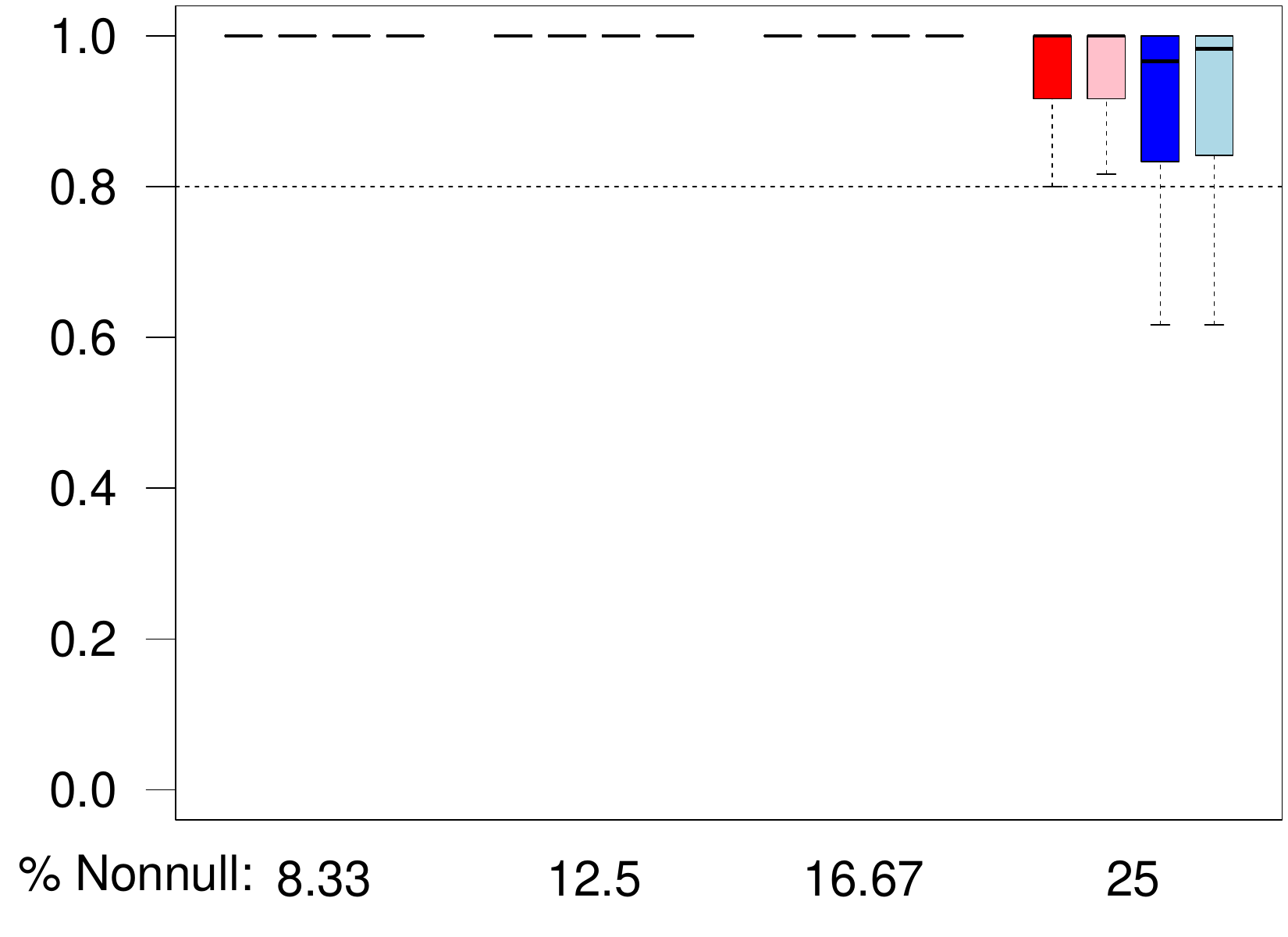}
		\includegraphics[width=0.23\textwidth]{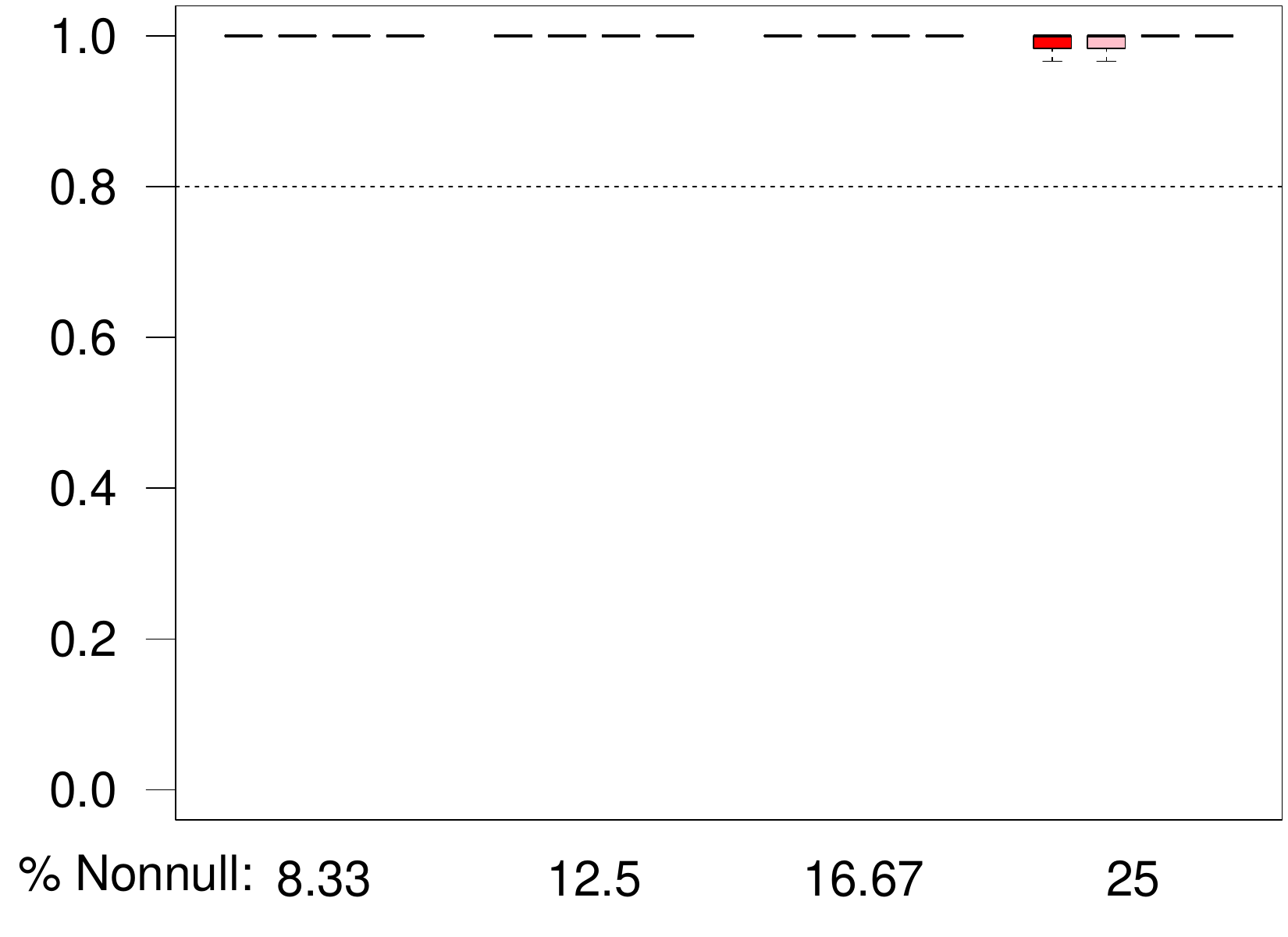}
		\includegraphics[width=0.23\textwidth]{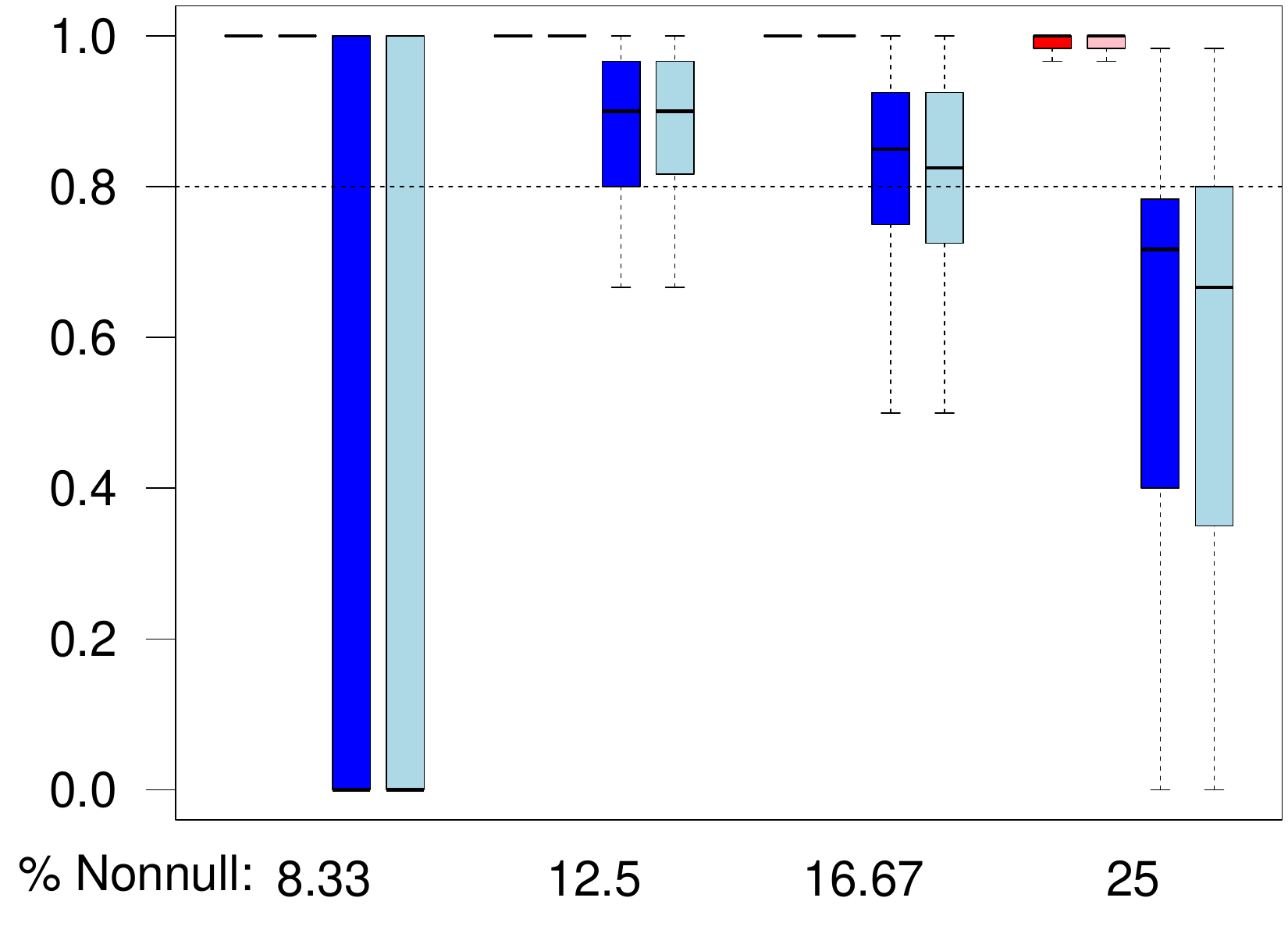}
		\includegraphics[width=0.23\textwidth]{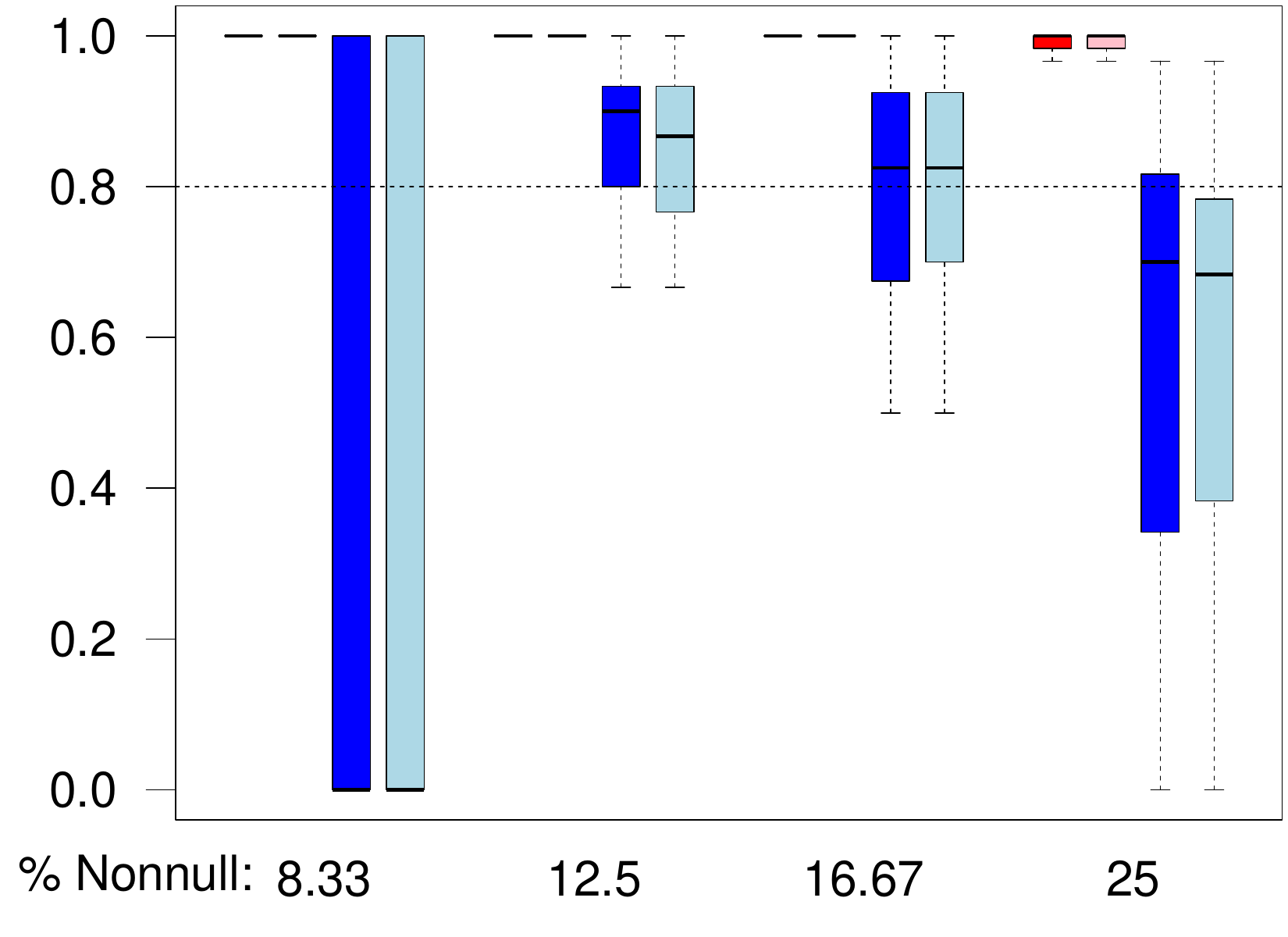}\\[-1cm]
	\end{tabular}
	\captionsetup{name=Figure} 
	\caption{Comparison of FDR and power for GDPM and Second-order Gaussian knockoffs. The left two columns show the results when there is only one cluster, while the right two columns show the results when the true data generating distribution is a mixture of two Gaussian distributions.}
	\label{fig:GDPMsim}
\end{table*}

Simulated datasets comprised 200 independent observations, with each observation featuring one continuous outcome and 240 real-valued predictors. The relationship between the outcome and predictors was modeled linearly as $y = \mathbf{X}^* \boldsymbol{\beta}^* + \epsilon$, with $\mathbf{X}^*$ representing the vector of non-null predictors and $\epsilon$ following a normal distribution, $\epsilon \sim \mathrm{No}(0,1)$. We explored four levels of sparsity in the datasets: $8.33\%$ (corresponding to 20 true features selected at random), $12.5\%$ (30), $16.67\%$ (40), and $25\%$ (60). For each sparsity scenario, 200 replicated datasets were created. To enhance statistical power, the lasso coefficient difference (LCD) was utilized as the knockoff statistic. In the context of Bayesian knockoff generator, the burn-in phase for the Markov Chain Monte Carlo (MCMC) chains was set at 1000 iterations.

\subsection{Comparision with Second Order Gaussian Knockoff Generator}
In this subsection, we examine four distinct true data generating distributions for the features: (1) the standard normal distribution, denoted as $\mathrm{No}(\mathbf{0},\mathbf{I})$; (2) a single normal distribution with a covariance matrix characterized by diagonal elements of 1 and off-diagonal elements adhering to an autoregressive structure with a correlation of $0.5$, represented as $\mathrm{No}(\mathbf{0},\mathrm{AR}(0.5))$; (3) a Gaussian mixture model with varying means but identity covariance matrices, expressed as $\ 0.5 \cdot \mathrm{No}(\mathbf{1},\mathbf{I}) + 0.5 \cdot \mathrm{No}(\mathbf{-1},\mathbf{I})$; and (4) a mixture of Gaussian distributions with an autoregressive covariance structure, $\ 0.5 \cdot \mathrm{No}(\mathbf{1},\mathrm{AR}(0.5)) + 0.5 \cdot \mathrm{No}(\mathbf{-1},\mathrm{AR}(0.5))$.

Signal strengths for non-null features were generated randomly from a Uniform distribution Unif$(1,2)$. The simulation outcomes, including empirical FDR and power estimates across different sparsity levels, target FDR settings, and covariate distributions, are summarized in Figure \ref{fig:GDPMsim} through boxplots. The first column of Figure \ref{fig:GDPMsim} reveals that under the standard normal distribution scenario, both the GDPM and the second-order Gaussian knockoff achieve FDR control close to the target level while exhibiting comparable power in identifying true features. Notably, in scenarios with a higher proportion of true non-null features (one-fourth of all features), the GDPM knockoff demonstrates a power advantage. For data drawn from a single normal distribution with feature correlation, the GDPM model performs better in FDR control and sustains high detection power.

When the features arise from a mixture of two Gaussian distributions, both the GDPM and the second-order Gaussian knockoff generator uphold FDR control. However, the Bayesian generator achieves FDR levels nearer to the target, with superior power across varying levels of feature sparsity and target FDR rates. This power benefit becomes more pronounced at lower target FDR thresholds (0.05) and across the spectrum of non-null feature proportions.

Overall, the Bayesian generator performs comparably to the second-order knockoff when predictors follow a Gaussian distribution, yet it provides more accurate FDR control and enhanced power in scenarios with more complex covariate distributions.

\subsection{Comparison with Finite Mixture Model}
To further assess the performance of the GDPM knockoff generator, we juxtapose it against the finite Gaussian mixture (FGM) knockoff approach proposed by \citet{GimeGhor19}. The true data generating distribution has four mixture components, and the mixture means were defined as $\boldsymbol{\mu}_1=(1,\dots,1,0,\dots,0)$, where the first half of the elements are $1$s, $\boldsymbol{\mu}_2=(0,\dots,0,1,\dots,1)$, with the first half as $0$s, $\boldsymbol{\mu}_3=-\boldsymbol{\mu}_1$, and $\boldsymbol{\mu}_4=-\boldsymbol{\mu}_2$. Two covariance scenarios were evaluated: one where the covariance is the identity matrix $\mathbf{I}$ for each component, and another featuring an autoregressive structure with a correlation of $0.5$ for each component.

To explore the effects of mis-specifying the number of clusters, the number of clusters for the FGM was deliberately set to $2$, instead of the actual number of $4.$ The signal strengths for non-null features were randomly drawn from a Unif$(0.5,1)$.

As depicted in the first row right column of Figure \ref{fig:FMM}, when the target FDR is set higher (0.2) and the features within each cluster are correlated, the FGM model significantly overestimates the type I error. Conversely, at lower target FDR settings, both methods effectively manage the false discovery rate. Nonetheless, the GDPM method consistently exhibits greater power compared to the FGM approach. The discrepancy in power is especially pronounced in scenarios lacking intra-cluster correlation, as opposed to those with feature correlation within clusters.
\setcounter{table}{2}
 \begin{table}[t!]
	\begin{tabular}{p{0.01cm}p{0.01cm}c}
	\multicolumn{3}{c}{\hspace{1.2cm} covariance matrix $\mathbf{I}$
		\hspace{0.8cm} covariance matrix $\mathrm{AR}(0.5)$}\\
	\multirow{2}{*}{\rotatebox[origin=c]{90}{\small Target FDR=0.2}} & 
	\rotatebox[origin=c]{90}{\hspace{2cm} \small FDR }&
	\includegraphics[width=0.2\textwidth]{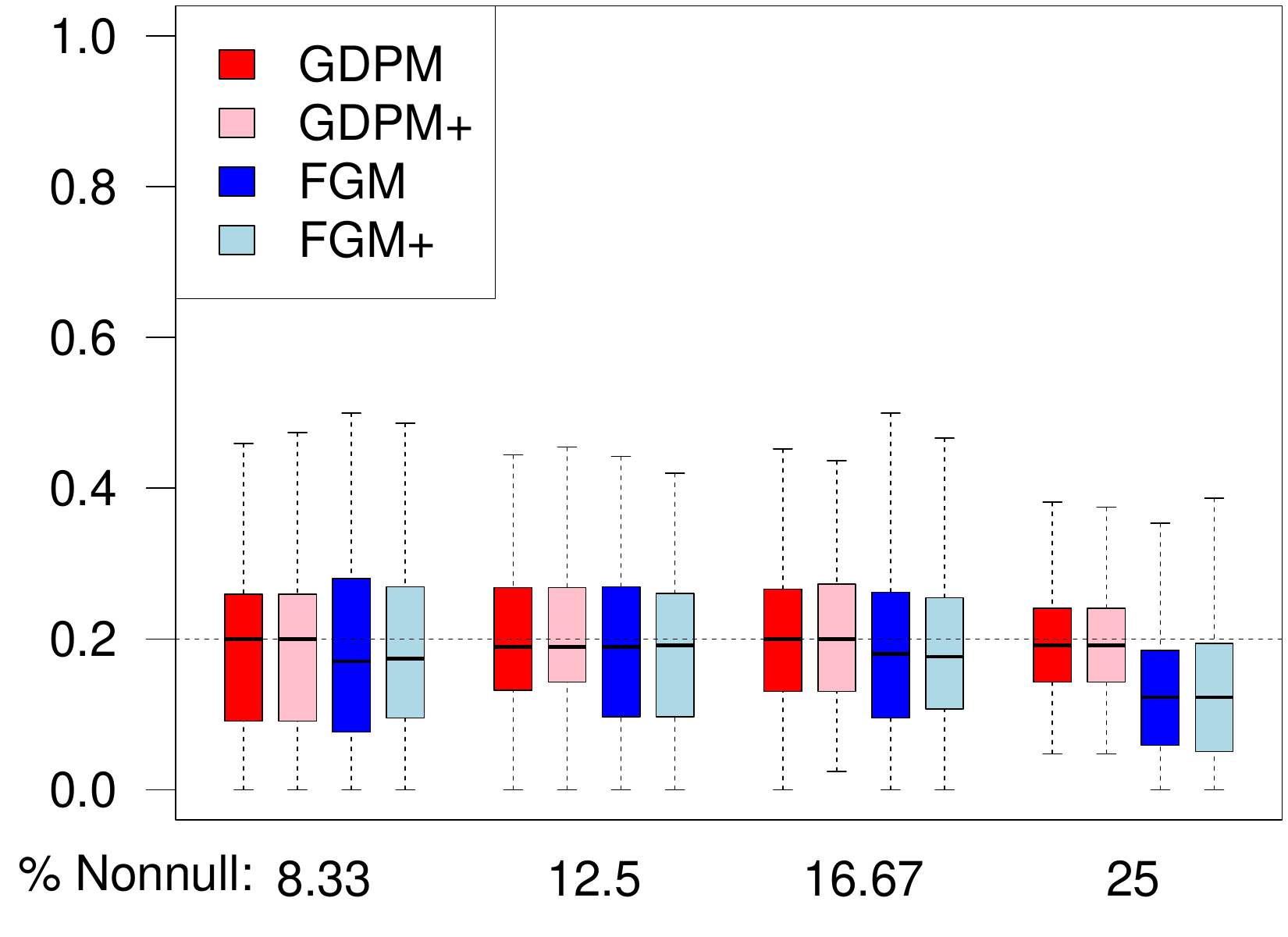}
	\includegraphics[width=0.2\textwidth]{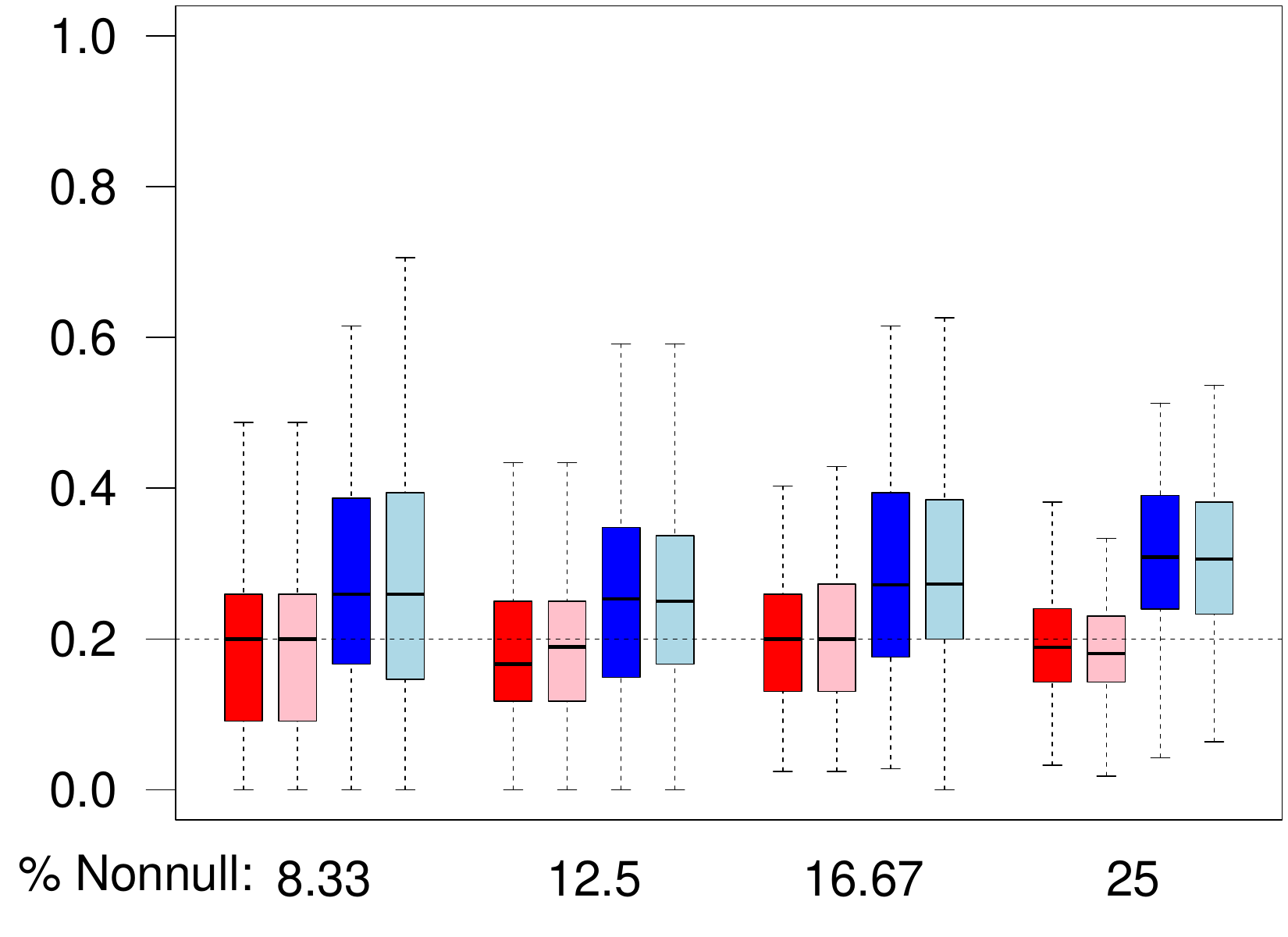}\\[-1cm]
	&\rotatebox[origin=c]{90}{\hspace{2cm}\small Power}&
	\includegraphics[width=0.2\textwidth]{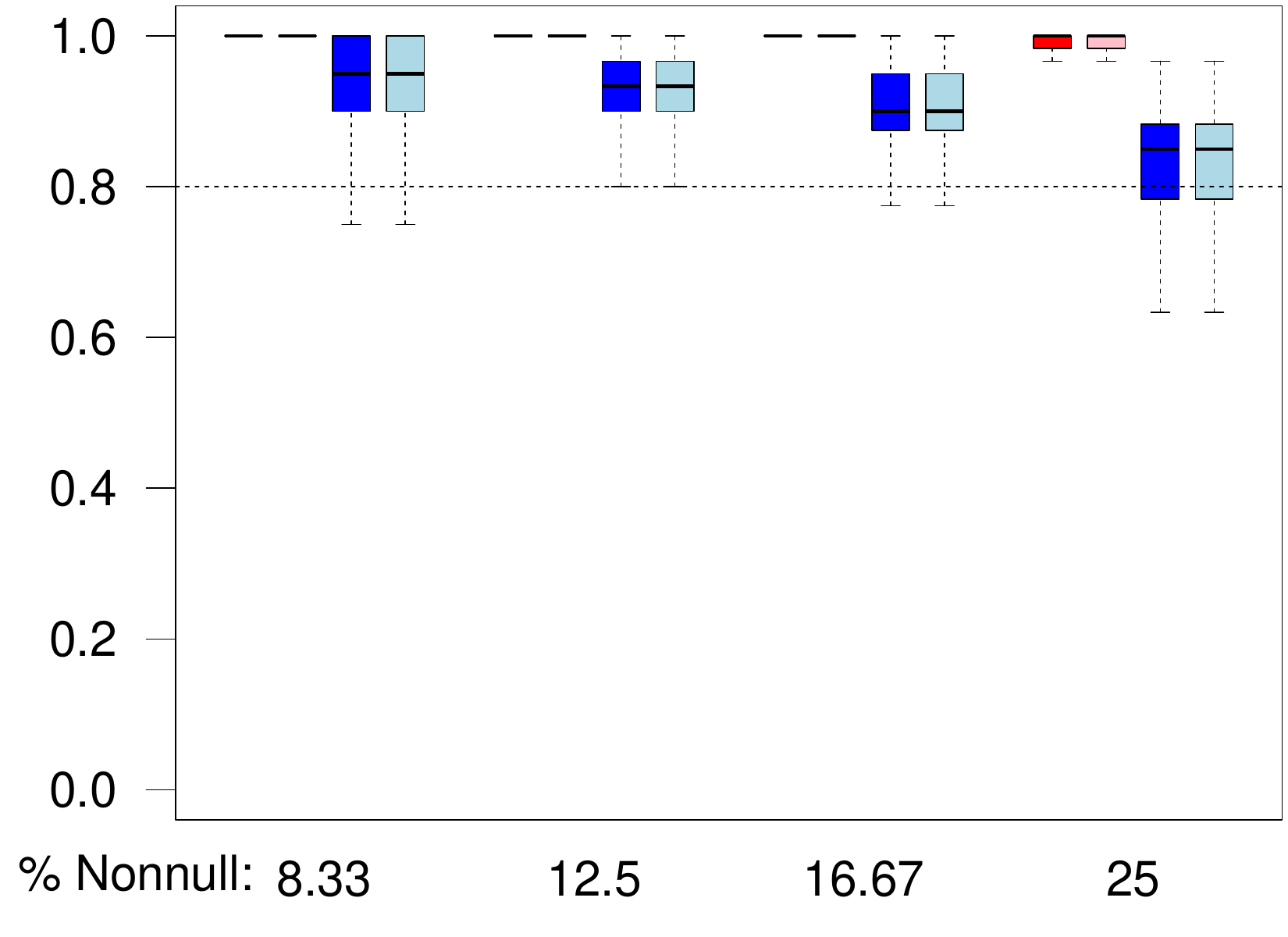}
	\includegraphics[width=0.2\textwidth]{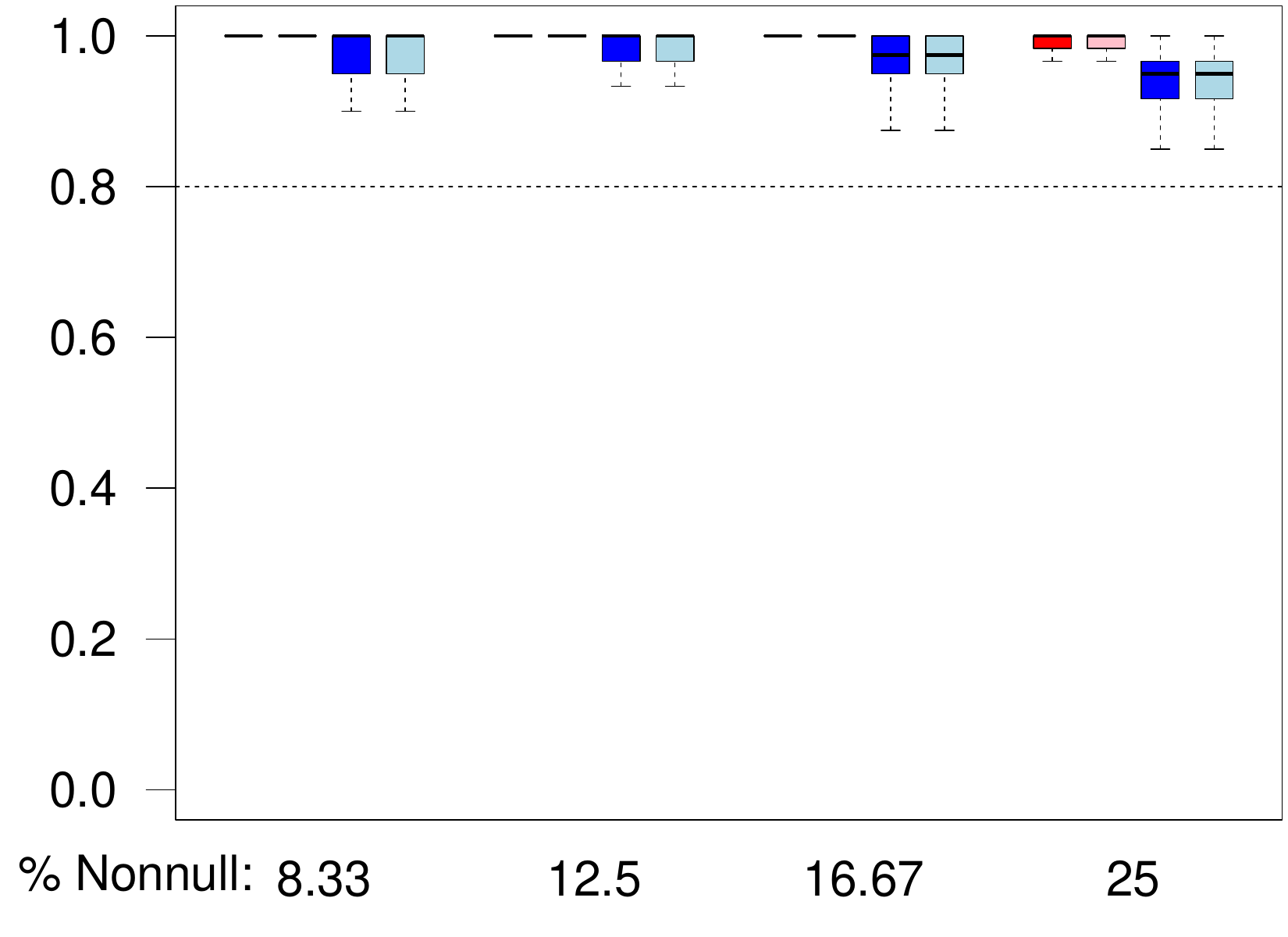}\\[-1cm]
\end{tabular}
\end{table}

\begin{table}[h!]\ContinuedFloat
	\begin{tabular}{p{0.01cm}p{0.01cm}c}
	\multicolumn{3}{c}{\hspace{1.2cm} covariance matrix $\mathbf{I}$
		\hspace{0.8cm} covariance matrix $\mathrm{AR}(0.5)$}\\
	\multirow{2}{*}{\rotatebox[origin=c]{90}{\small Target FDR=0.1}} & 
	\rotatebox[origin=c]{90}{\hspace{2cm} \small FDR }&
	\includegraphics[width=0.2\textwidth]{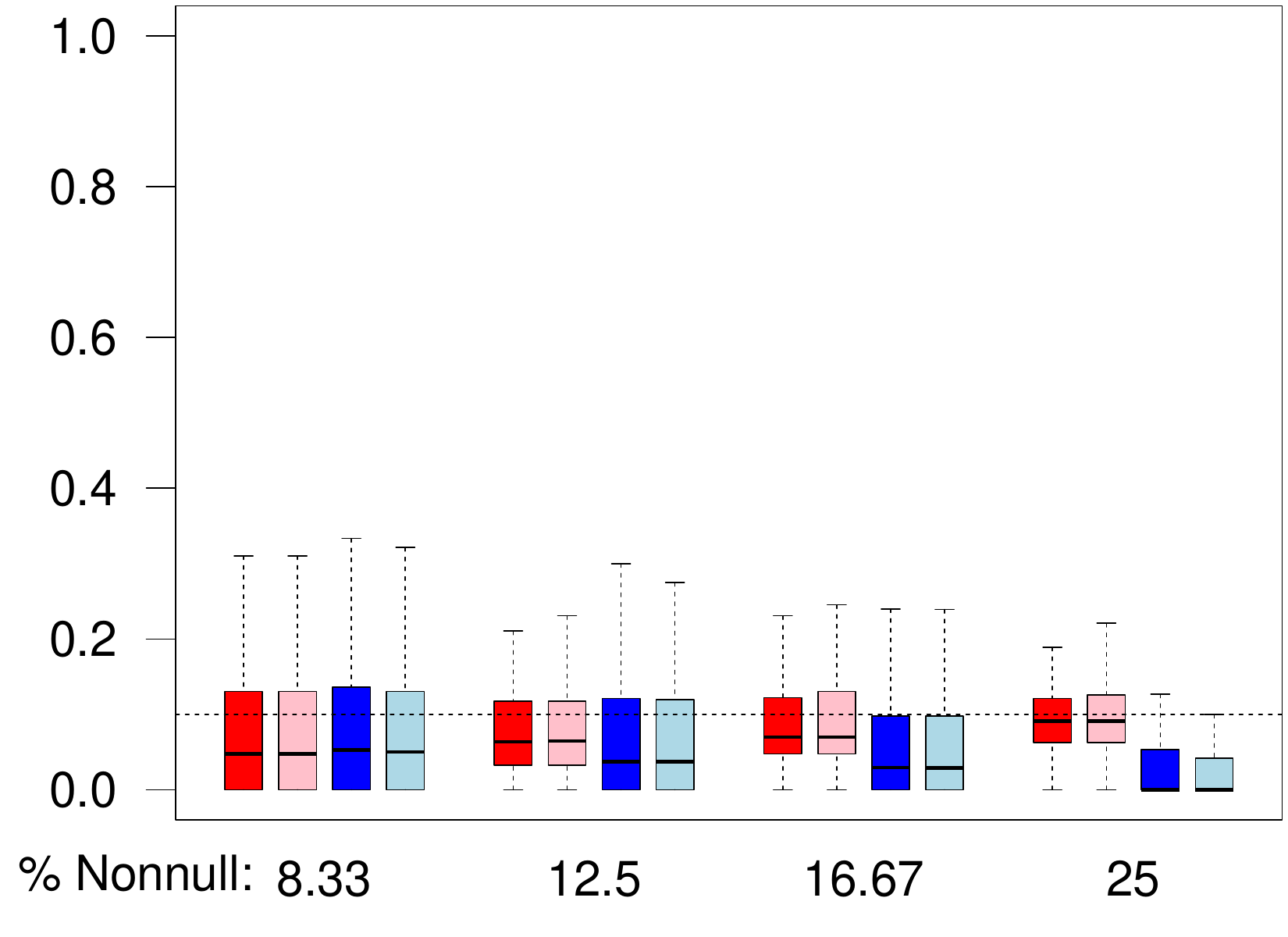}
		\includegraphics[width=0.2\textwidth]{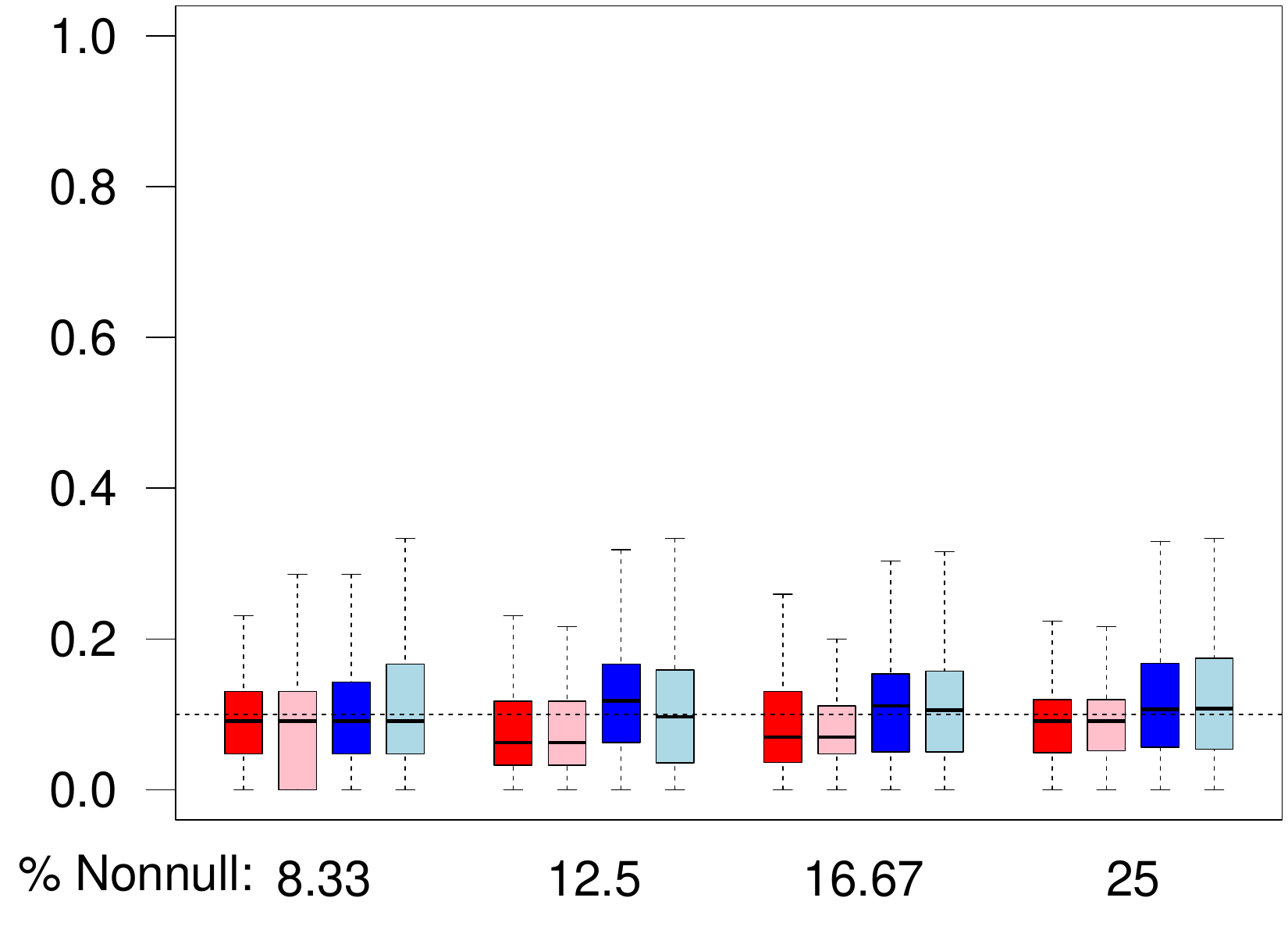}\\[-1cm]
	&\rotatebox[origin=c]{90}{\hspace{2cm}\small Power}&
	\includegraphics[width=0.2\textwidth]{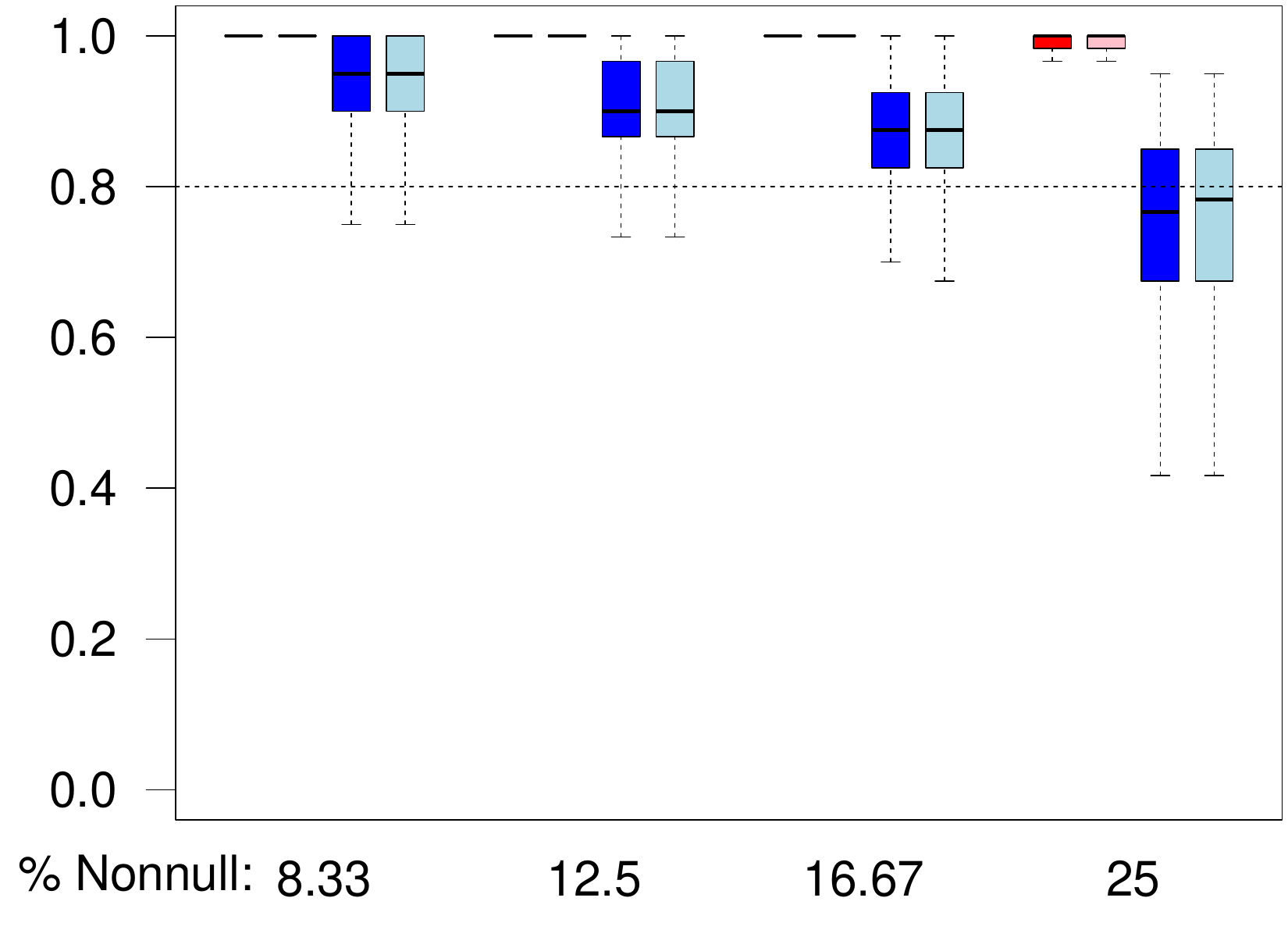}
	\includegraphics[width=0.2\textwidth]{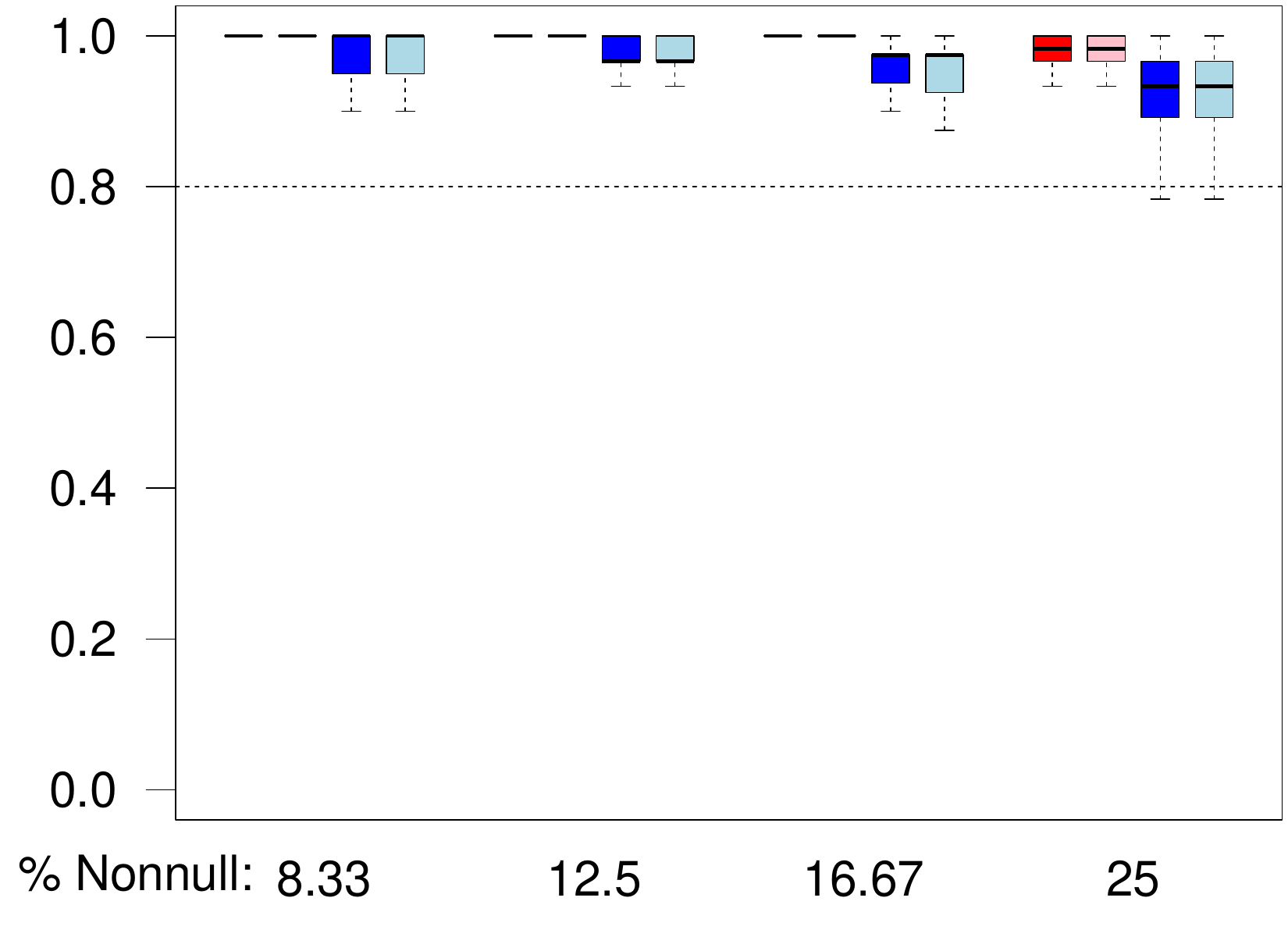}\\[-1cm]
\end{tabular}
\end{table}	

\begin{table}[h!]\ContinuedFloat
	\begin{tabular}{p{0.01cm}p{0.01cm}c}
	\multicolumn{3}{c}{\hspace{1.2cm} covariance matrix $\mathbf{I}$
		\hspace{0.8cm} covariance matrix $\mathrm{AR}(0.5)$}\\
	\multirow{2}{*}{\rotatebox[origin=c]{90}{\small Target FDR=0.05}} & 
	\rotatebox[origin=c]{90}{\hspace{2cm} \small FDR }&
	\includegraphics[width=0.2\textwidth]{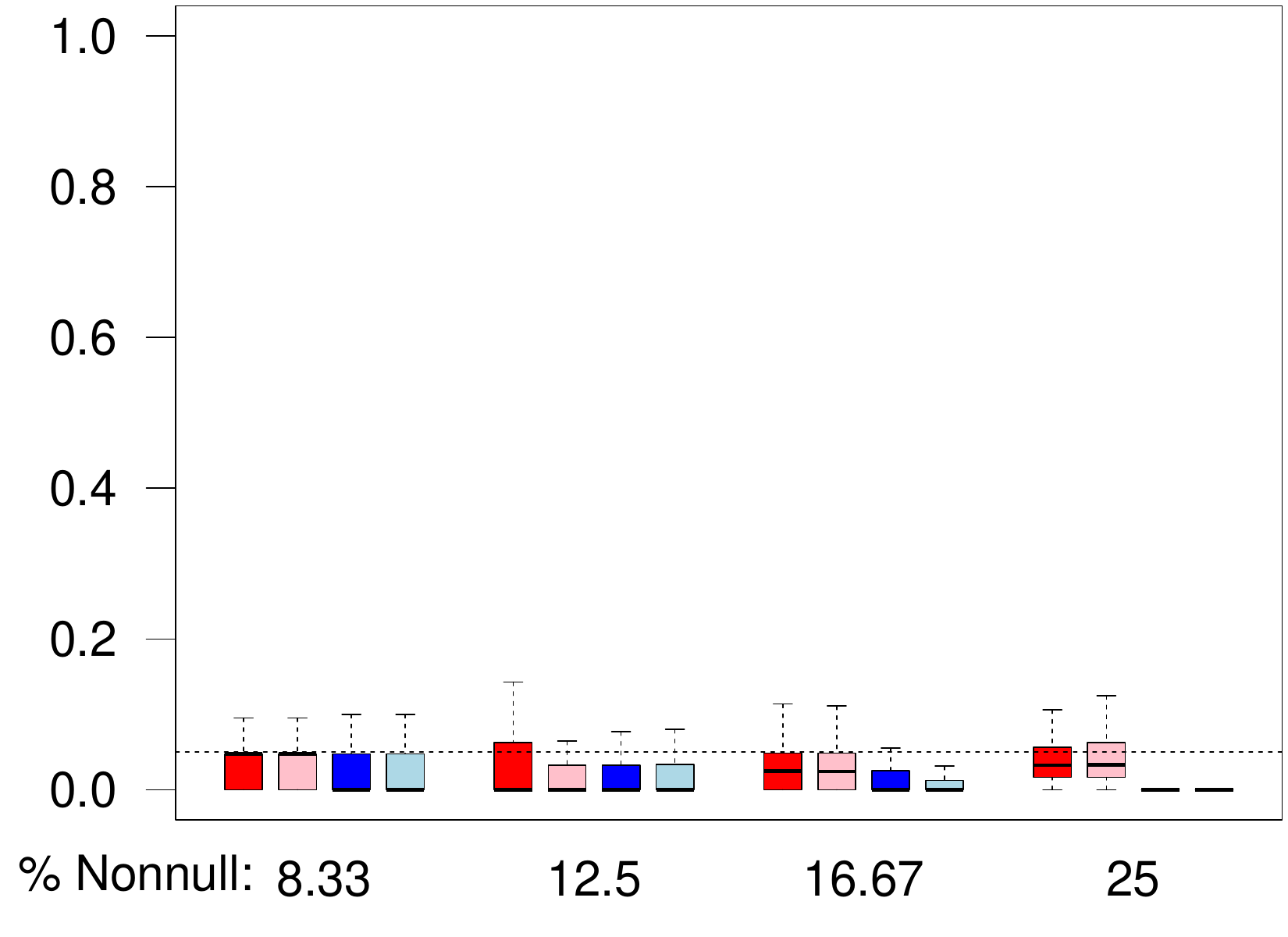}
		\includegraphics[width=0.2\textwidth]{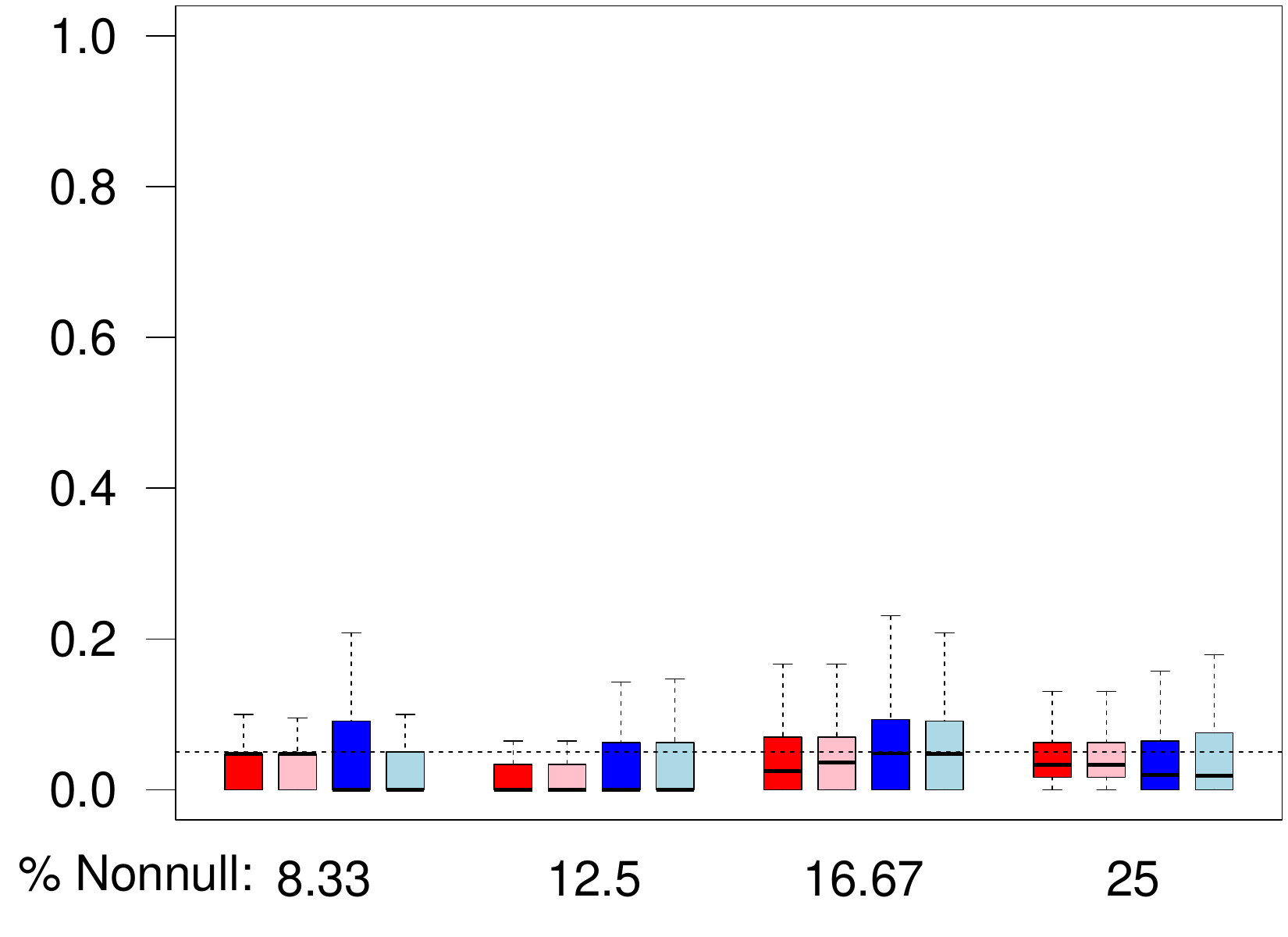}\\[-1cm]
	&\rotatebox[origin=c]{90}{\hspace{2cm}\small Power}&
	\includegraphics[width=0.2\textwidth]{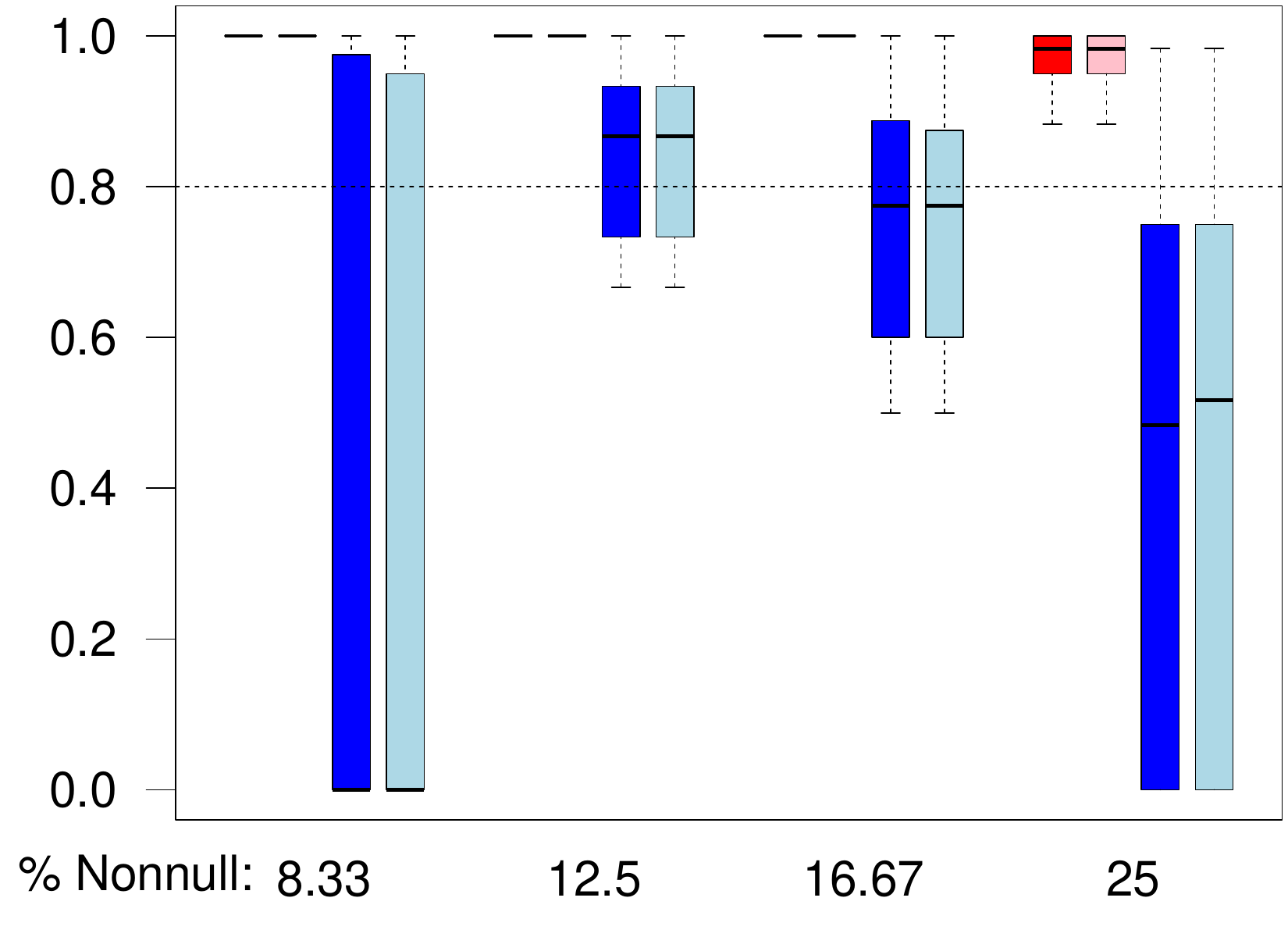}
	\includegraphics[width=0.2\textwidth]{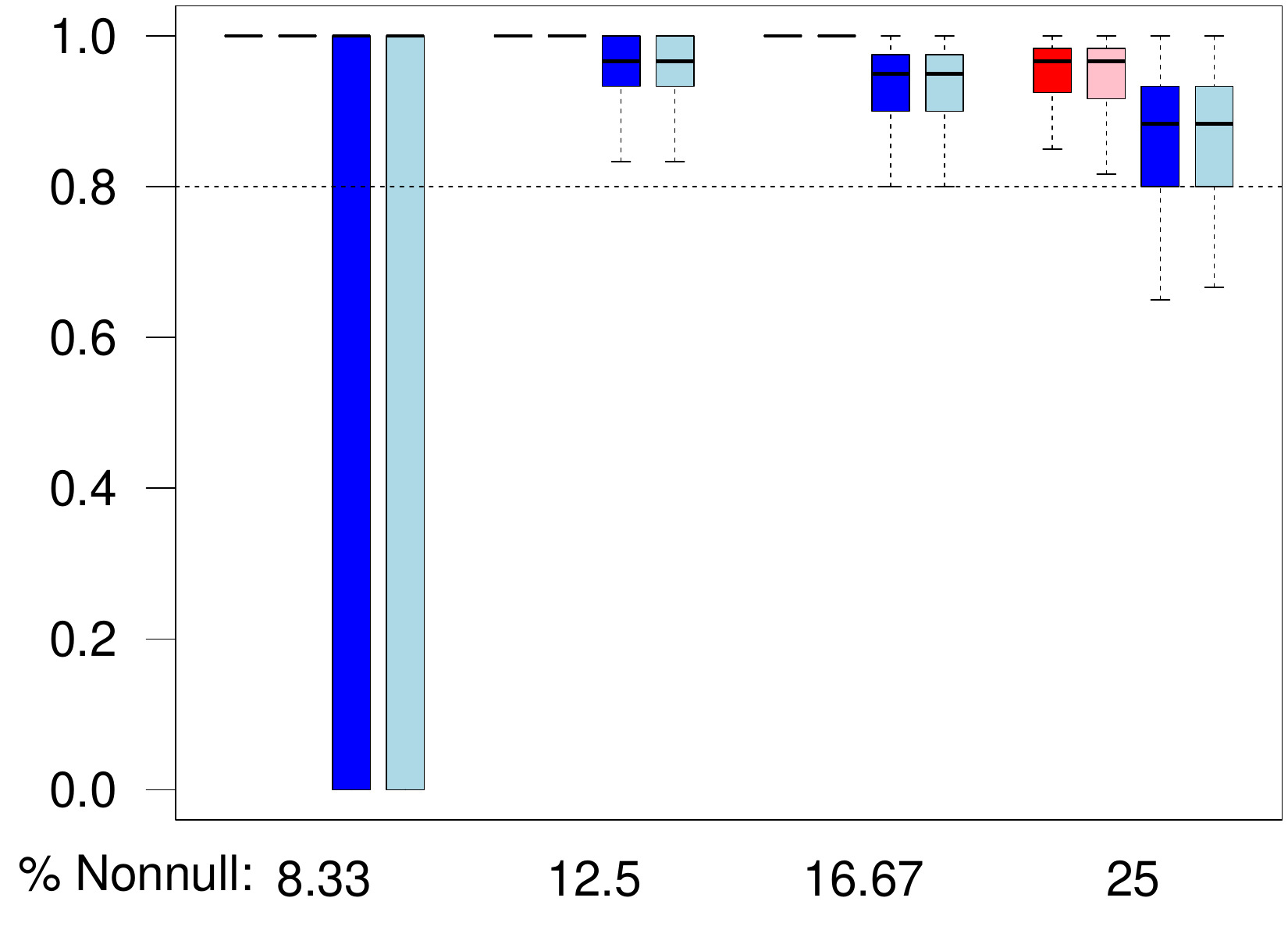}\\[-1cm]
\end{tabular}
	\captionsetup{name=Figure} 
	\caption{Comparison of FDR and power for GDPM and FGM knockoffs. The left column shows the results when the covariance in each cluster is identity matrix, while the right column shows the results when covariance matrix in each cluster an autoregressive structure.}

	\label{fig:FMM}
\end{table}

\subsection{Computing Time}
We have implemented GDPM model and knockoff generation in RCpp to boost computational speed. For the simulated data with 2 underlying clusters described in Section 4, a computer with an Intel Xeon E5-4610 v2 2.3GHz processor takes 8 minutes to run 200 MCMC iterations for the GDPM model.

\section{SELECTING GENETIC PREDICTORS OF SURVIVAL IN OVARIAN CANCER PATIENTS}
We evaluated the prognostic value of mRNA gene expression levels for overall survival in 559 ovarian cancer patients using The Cancer Genome Atlas (TCGA) data. The data include 18632 gene expression features and times of death or last follow-up for these patients. The expression levels were log-transformed and screened to retain only the most diverse features for selection. Genes with standard deviation greater than 0.15 were retained, producing a set of 576 candidate features for knockoff selection. 

The GDPM model was used to generate knockoff features for this set with the recommended prior in Section 2. The importance statistic used for knockoff selection was the LSM statistic from a Cox lasso model of the overall survival time, assuming random censoring. Eleven genes were selected by this procedure:
CXCL13, B3GALT1, D4S234E, ZYG11A, LEMD1, OVGP1, CST9, UBD, NRG4, CCDC80, and FLJ21963. The estimated hazard ratios (HRs) from a reduced Cox model fitted using only the selected features are shown in Table \ref{mrnaTable}. Previously, increasing UBD, CXCL13, and D4S234E were identified as having protective effects (HR $< 1$) on overall survival in a nine-gene panel constructed by \citet{DingDong20} from the TCGA database using a Cox lasso model. \citet{GaoLiu17} found OVGP1 to be protective with regard to overall survival in an analysis that found associations of 11 genes with survival using univariate log-rank tests of each candidate gene. 
Because differing screening methods and/or analysis methods were used in each study to obtain its final panel of genes, strong agreement is not expected between our set of selected genes and theirs. But, it is reassuring that some genes are shared across studies with estimated effects in the same direction.
\setcounter{table}{1}
\setcounter{table}{0}
\begin{table}[t!] 
	\centering
	\caption{Estimated hazard ratios from reduced Cox model with selected genes from TCGA ovarian cancer dataset. Features identified by previous literature are shown in italicized font.}
	\begin{tabular}{cc|cc}\hline
		\textbf{GENE}&\textbf{HAZARD } & \textbf{GENE}&\textbf{HAZARD} \\
		&\textbf{RATIO} & &\textbf{RATIO} \\\hline
		\textit{CXCL13}& $0.385$  & CST9& $0.551$\\
		B3GALT1& $1.742$ & \textit{UBD}& $0.537$\\
		\textit{D4S234E}& $0.342$  &NRG4& $0.331$\\
		ZYG11A& $1.693$ &  CCDC80& $2.708$\\
		LEMD1& $0.722$ &  FLJ21963& $2.336$\\
		\textit{OVGP1}& $0.586$ &&\\\hline
	\end{tabular}
	
	\label{mrnaTable}
\end{table}

\section{DISCUSSION}
We proposed a Dirichlet process mixture model for knockoff generation of feature vectors, providing additional flexibility for modeling complex distributions. The method is shown to identify important features with good FDR control and high power through simulations and real data application. Though we rely on MCMC approaches to produce knockoff copies for the simulations and application in this paper, other posterior inference methods such as variational inference can be used and should be considered. These alternatives may be especially valuable for high or ultrahigh dimensional feature sets where exact inference is computationally expensive. In this paper, the method developed is for continuous predictors; for data with categorical predictors, we propose to use \citet{PolsScot13}'s P\'{o}lya-Gamma augmented Gibbs sampler to sample the posterior. Our group plans to provide an Rcpp-based efficient implementation of the GPDM model with P\'{o}lya-Gamma augmented sampler for categorical covariates for moderate to large datasets in the near future.

Our methods focus on improving the quality of knockoff generation for complex data contexts. Another direction in the knockoff literature that has gained less attention is the choice of importance statistic for filtering. \citet{CandFan18} suggested considering Bayesian models when computing importance statistics and noted that the inclusion of accurate prior information in such models can boost the power over the LCD. We plan to study the use of Bayesian importance statistics in conjunction with our Bayesian knockoff generators to further enhance knockoff filtering.

\section{APPENDIX}
In this appendix, we provide detailed proof of Theorem 1 presented in the main text.


We can write
\begin{align*}
	\widehat{KL}_j = \sum_{i=1}^n \log R_{ij} - \log \widetilde{R}_{ij},
\end{align*}
where 
\begin{align*}
	R_{ij} &= \frac{p_j^*(\mathbf{X}_{ij} | \mathbf{X}_{i, -j}) }{p_j^m(\mathbf{X}_{ij} | \mathbf{X}_{i, -j})} = \frac{p_X^*(\mathbf{X}_{ij}, \mathbf{X}_{i, -j}) q_j^m(\mathbf{X}_{i, -j}) }{p_X^m(\mathbf{X}_{ij}, \mathbf{X}_{i, -j}) q_j^*( \mathbf{X}_{i, -j})}, \\
	\widetilde{R}_{ij} &= \frac{p_j^*(\widetilde{\mathbf{X}}_{ij} | \mathbf{X}_{i, -j}) }{p_j^m(\widetilde{\mathbf{X}}_{ij} | \mathbf{X}_{i, -j})} = \frac{p_X^*(\widetilde{\mathbf{X}}_{ij}, \mathbf{X}_{i, -j}) q_j^m(\mathbf{X}_{i, -j}) }{p_X^m(\widetilde{\mathbf{X}}_{ij}, \mathbf{X}_{i, -j}) q_j^*( \mathbf{X}_{i, -j})},
\end{align*}
$p_X^m$ is the estimated joint distribution of the features based on a GDPM model with $m$ observations, and $q_j^*$ and $q_j^m$ are the true and estimated marginal distributions for $(X_1,X_2, \dots, X_{j-1}, X_{j+1}, \dots, X_p)$ from this model.

By Theorem 3 of \cite{WuGhos10}, the posterior distribution $P_X^m$ of the GDPM, given $m$ data points, is L1-consistent for the true data distribution $P_X^*$. Then 
\begin{align*}
	\lim_{m \to \infty} \int_{\mathbb{R}^p} |p_X^m(\mathbf{x}) - p_X^*(\mathbf{x})| d\mathbf{x} = 0.
\end{align*}

Let $(\Omega, \mathcal{F}, P)$ denote the probability space upon which the feature vector $\mathbf{X}$ is defined. Then the induced probability measure $P_X = P \circ X^{-1}$ is such that $(\mathbb{R}^p, \mathcal{B}, P_X)$ is a probability space, where $\mathcal{B}$ is the collection of Borel sets of $\mathbb{R}^p$. 

For any $m$, the functions $p_X^m, p_X^*: \mathbb{R}^p \to \mathbb{R}$ are measurable functions mapping from elements of $(\mathbb{R}^p, \mathcal{B})$ to $(\mathbb{R}, \lambda)$, where $\lambda$ denotes Lebesgue measure on $\mathbb{R}$. Then the same is true for $|p_X^m - p_X^*|$. For any $\delta > 0$, define $A_\delta = \{\mathbf{x} \in \mathbb{R}^p: |p_X^m(\mathbf{x}) - p_X^*(\mathbf{x})| > \delta \} = \big[|p_X^m - p_X^*|\big]^{-1} (\delta,\infty)$. Then $A_\delta$ is the preimage of an open interval with respect to a measurable function, so is a $P_X$-measurable set.

By L1 consistency of $p_X^m$ for $p_X^*$, for any $\epsilon > 0$ there exists $M_\epsilon$ such that $m \ge M_\epsilon$ implies
\begin{align*}
	\epsilon \cdot \delta > \int_{\mathbb{R}^p} |p_X^m(\mathbf{x}) - p_X^*(\mathbf{x})| d\mathbf{x}.
\end{align*}
Since $|p_X^m - p_X^*|$ is nonnegative, then $m \ge M_\epsilon$ implies
\begin{align*}
	\epsilon \cdot \delta >& \int_{\mathbf{x} \in A_\delta} |p_X^m(\mathbf{x}) - p_X^*(\mathbf{x})| d\mathbf{x} \\
	&+ \int_{\mathbf{x} \in A_\delta^c} |p_X^m(\mathbf{x}) - p_X^*(\mathbf{x})| d\mathbf{x} \\
	\ge &\int_{\mathbf{x} \in A_\delta} \delta d\mathbf{x} + 0 \\
	=& \delta \cdot P_X(A_\delta),
\end{align*}
or $\epsilon > P_X(A_\delta)$. This implies that for any $\delta > 0,$ $\lim_{m \to \infty} P_X (\{\mathbf{x} \in \mathbb{R}^p: |p_X^m(\mathbf{x}) - p_X^*(\mathbf{x})| > \delta \}) = 0.$ Thus, $p_X^m \xrightarrow{P_X} p_X^*$. \\

We can write the marginal densities $q_j^m$ and $q_j^*$ as $q_j^m(\mathbf{X_{-j}}) = \int_{x_j \in \mathbb{R}} p_X^m(\mathbf{X}) dx_j$ and $q_j^*(\mathbf{X_{-j}}) = \int_{x_j \in \mathbb{R}} p_X^*(\mathbf{X}) dx_j$, where $\mathbf{X_{-j}} = (X_1,X_2, \dots, X_{j-1}, X_{j+1}, \dots, X_p)$. Since $p_X^m \xrightarrow{P_X} p_X^*$ and $p_X^*$ is assumed to be bounded by condition B1 of \cite{WuGhos10}, the Dominated Convergence Theorem implies that $q_j^m \xrightarrow{P_X} q_j^*$ for each $j$. Also, the $\log$ function is continuous on $(0,\infty)$. By the Continuous Mapping Theorem, $\log p_X^m \xrightarrow{P_X} \log p_X^*$ and $\log q_X^m \xrightarrow{P_X} \log q_X^*$. Thus, $\log p_X^* - \log p_X^m + \log q_j^m - \log q_j^* \xrightarrow{P_X} 0$. Then for any $\eta > 0$ there exists $K_\eta^j$ such that $m \ge K_\eta^j$ implies 
\begin{align*}
	&P_X\left(|\log R_{ij}| > \frac{\eta}{2n}\right) \\
	=& P_X\left(|\log p_X^*(\mathbf{X}_{ij}, \mathbf{X}_{i, -j}) - \log p_X^m(\mathbf{X}_{ij}, \mathbf{X}_{i, -j})\right.\\
	&\left. + \log q_j^m(\mathbf{X}_{i, -j}) - \log q_j^*( \mathbf{X}_{i, -j})| > \frac{\eta}{2n}\right) \\
	<& \frac{\eta}{2n}.
\end{align*}
Similarly, $P_X\left(|\log \widetilde{R}_{ij}| > \frac{\eta}{2n}\right) < \frac{\eta}{2n}$ if $m \ge K_\eta^j$. By the triangle inequality, it follows that $P_X(\widehat{KL}_j > \eta) < \eta$ when $m \ge K_\eta^j$. Letting $K_\eta = \max_j K_\eta^j,$ we have $P_X(\max_j \widehat{KL}_j > \eta) < \eta$ when $m \ge K_\eta$.

Finally, for any $\epsilon > 0$, let $\eta = \min\{\epsilon/2,\log(1+\epsilon/(2q)\}$ and set $M_\epsilon = K_\eta$. Using the FDR bound from Theorem 1 of \cite{BarbCand20}, if $m \ge M_\epsilon,$ then
\begin{align*}
	FDR &\le q e^\eta + P\left(\max_{j \in H_0} \widehat{KL}_j > \eta \right)\\
	&\le q\left(1+\frac{\epsilon}{2q}\right) + \eta \\
	&\le q + \frac{\epsilon}{2} + \frac{\epsilon}{2} \\
	&= q + \epsilon.
\end{align*}

By choosing $\epsilon$ to be as small as desired, the FDR/mFDR can be bounded arbitrarily close to $q$ under Knockoff+/Knockoff filtering, provided that $m$ is sufficiently large. \qedsymbol

\end{document}